# Search for like-sign dileptons plus two jets signal in the framework of the manifest left-right symmetric model

**Aviad Roitgrund and Gad Eilam**[1]

*Department of Physics, Technion — Israel Institute of Technology, Haifa 32000, Israel*

*E-mail:* aviadroi@gmail.com

Abstract: The left-right symmetric model may present evidence of new physics at the era of the Large Hadron Collider (LHC). We use its framework to investigate the lepton number violating process $pp \to e^{\pm}e^{\pm}jj + X$ at the 14 TeV LHC. We show that for an integrated luminosity of 300 fb$^{-1}$, the discovery contour of the right-handed boson $W_R$ and the right-handed electron neutrino $N_e$ as a result of the $pp \to W_R^{\pm} \to e^{\pm}N \to e^{\pm}e^{\pm}jj$ process can be expanded upon considering an additional channel mediated by the right-handed doubly charged Higgs $\delta_R^{\pm\pm}$, i.e. $pp \to W_R^{\pm} \to W_R^{\mp *}\delta_R^{\pm\pm *} \to e^{\pm}e^{\pm}jj$.

[1] Now deceased.

## Contents

## 1 Introduction

The shortcomings of the Standard Model (SM) are the motivation to search for signals of new physics beyond it. This is the main goal of the Large Hadron Collider (LHC). The SM is incapable of explaining a number of fundamental issues, such as the hierarchy problem (resulting from the large difference between the weak force and the gravitational force), the dark matter problem and the number of families in the quark and lepton sector. It is, therefore, reasonable to assume that new physics beyond the SM will be discovered in the coming years. Among the possible attractive platforms for new physics are left-right symmetric models (LRSM) [1–7].

The LRSM addresses two specific deficiencies in the SM: (i) Parity violation in the weak interactions, and (ii) non-zero neutrino masses implied by the experimental evidence of neutrino oscillation [8]. In particular, the left-right symmetry which underlies LRSM restores Parity symmetry at energies appreciably higher than the electroweak scale, resulting in the addition of three new gauge boson fields, $W_{R1,2,3}$. Furthermore, in LRSM the neutrinos are massive and their nature (i.e., whether they are of Majorana or Dirac type) depends on the details of the LRSM.

Early constructions of the LRSM comprise a Higgs sector with a Higgs bidoublet and two Higgs doublets [1–3]. In such a setup, the neutrinos are of Dirac type and no natural explanation for their small masses is provided. A later version, the manifest (or quasi-manifest — see below) LRSM, incorporates a Higgs bidoublet and two Higgs triplets, and implies the existence of Majorana type neutrinos [4–7]. This later version provides a natural setup for the smallness of neutrino masses, relating their mass scale to the large left-right symmetry breaking scale through the see-saw mechanism [9–11].

In this work we examine a process which, if detected at the LHC, will be able to provide direct evidence to the correctness of the LRSM with regard to two primary points: the left-right symmetry breaking scale and the see-saw mechanism. The process, which is lepton number violating (LNV), is

$$pp \to l^{\pm} l^{\pm} + 2j + X. \tag{1.1}$$

This process has a unique signal of two same sign leptons (like-sign dileptons — LSD) and two jets (LSD + 2 jets), with no missing energy. It has two classes of channels: the first class consists of channels which comprise the heavy right-handed (RH) boson $W_R$ and the RH Majorana neutrino $N$ [12, 13],[1] and the second class consists of channels which comprise $W_R$ and the RH doubly charged Higgs $\delta_R^{\pm\pm}$ [14–16]. The signal of the process, if observed, may allow one to

1. detect the RH gauge boson $W_R$,

2. trace the see-saw mechanism and the Majorana nature of the neutrino by detecting a heavy RH neutrino — which is associated to the spontaneous left-right symmetry breaking scale by its heavy mass,

3. establish the existence of the charged Higgs bosons and further confirm the Higgs triplet nature.

In this work we investigate the above process at the (14 TeV) LHC within the framework of the manifest LRSM. Former studies of the LSD + 2 jets signal at the LHC (within the framework of the LRSM) focused on a Drell-Yan production of a RH neutrino and a lepton via the $W$ boson, followed by the neutrino decay to a second same sign lepton and two jets [17–33]:

$$pp \to W_{L,R}^{\pm} \to l^{\pm} N \to l^{\pm} l^{\pm} jj. \tag{1.2}$$

We extend these studies by including all the possible diagrams and searching for the dominant contributions among the possible LRSM amplitudes of the process. We find that an additional channel, mediated by the RH gauge boson and the doubly charged Higgs,

$$pp \to W_R^{\pm} \to W_R^{\mp *} \delta_R^{\pm\pm *} \to l^{\pm} l^{\pm} jj, \tag{1.3}$$

can significantly contribute to the LSD + 2 jets signal when considering the reach of the LHC at $\sqrt{s} = 14\,\text{TeV}$.

The work is organized as follows. In section 2 we describe the manifest LRSM Lagrangian structure. In section 3 we briefly mention constraints on the parameter space which are relevant to the study of the LSD + 2 jets phenomenology. In section 4 we investigate the process. We find and evaluate the cross sections of its dominant channels, and reconstruct the heavy gauge boson $W_R$, the RH electron neutrino $N_e$ and the doubly charged Higgs $\delta_R^{\pm\pm}$ from the relevant kinematic variables of signal and background events. We then plot the discovery potential (associated with the process) of these particles at the LHC ($\sqrt{s} = 14\,\text{TeV}$). We summarize in section 5.

[1] The neutrinoless double $\beta$ decay (discussed in ref. [13]), although applicable to the LRSM, was not proposed within its context.

## 2 General model description

The LRSM is based on the gauge group $\text{SU}(3)_C \times \text{SU}(2)_L \times \text{SU}(2)_R \times \text{U}(1)_{B-L}$. All the fermion fields in the model are assigned in doublets, including the RH fermions which transform as doublets under the new symmetry of the model, $\text{SU}(2)_R$. In addition to the fermion fields, seven gauge fields, $\vec{W}_{L,R}$ and $B$ (corresponding to the groups $\text{SU}(2)_{L,R}$ and $\text{U}(1)_{B-L}$ respectively) and eight gluon fields $G_a\,(a = 1 \ldots 8)$ are introduced in order to obtain gauge invariance. The appropriate coupling constants of the $G^a_\mu$, $\vec{W}_{L,R\,\mu}$ and the $B_\mu$ fields are $g_s$, $g_{L,R}$ and $g' = g_{B-L}$, respectively. The scalar content of the model includes three Higgs multiplets: a bidoublet (denoted as $\phi$), a RH and a left-handed (LH) triplet (denoted as $\mathbf{\Delta}_L$ and $\mathbf{\Delta}_R$, respectively). The covariant derivatives of the multiplets are conventionally given in the adjoint representation, so that the triplets are also converted to the adjoint representation. Thus, the $2 \times 2$ bidoublet-equivalent field matrices $\Delta_{L,R} = \frac{1}{\sqrt{2}}\vec{\sigma} \cdot \mathbf{\Delta}_{L,R}$ are introduced (the three-vector $\vec{\sigma}$ contains the Pauli matrices as components). The model field content is given in table 1.

The requirement that the Lagrangian is invariant under the left-right symmetry ($\psi$ represents the $L$ and the $Q$ fermion doublets, see table 1)

$$\psi_L \leftrightarrow \psi_R, \quad \vec{W}_L \leftrightarrow \vec{W}_R, \quad \Delta_L \leftrightarrow \Delta_R, \quad \phi \leftrightarrow \phi^\dagger, \tag{2.1}$$

leads to

$$g_L = g_R. \tag{2.2}$$

As spontaneous symmetry breaking (SSB) occurs, the above Higgs multiplets break the left-right symmetry into the $\text{U}(1)_Q$ observed symmetry, where the electromagnetic charge $Q$ is defined by the modified Gell-Mann-Nishijima formula [34]

$$Q = I_{3L} + I_{3R} + \frac{B-L}{2}. \tag{2.3}$$

At the first stage of the SSB the RH Higgs triplet, $\Delta_R$, acquires a VEV $v_R$:

$$\langle \Delta_R \rangle = \frac{1}{\sqrt{2}} \begin{pmatrix} 0 & 0 \\ v_R & 0 \end{pmatrix}. \tag{2.4}$$

This stage occurs at an energy scale which is much larger than the electroweak scale. At the second stage the bidoublet acquires a VEV

$$\langle \phi \rangle = \frac{1}{\sqrt{2}} \begin{pmatrix} k_1 & 0 \\ 0 & k_2 \end{pmatrix}. \tag{2.5}$$

| Model fields | Content | $SU(3)_C$ | $\times$ $SU(2)_L$ | $\times$ $SU(2)_R$ | $\times$ $U(1)_{B-L}$ |
|---|---|---|---|---|---|
| $L_{iL}$ | $\begin{pmatrix} \nu'_i \\ l'_i \end{pmatrix}_L$ | 1 | 2 | 1 | -1 |
| $L_{iR}$ | $\begin{pmatrix} \nu'_i \\ l'_i \end{pmatrix}_R$ | 1 | 1 | 2 | -1 |
| $Q_{iL}$ | $\begin{pmatrix} u'_i \\ d'_i \end{pmatrix}_L$ | 3 | 2 | 1 | $\frac{1}{3}$ |
| $Q_{iR}$ | $\begin{pmatrix} u'_i \\ d'_i \end{pmatrix}_R$ | 3 | 1 | 2 | $\frac{1}{3}$ |
| $W_L$ | $W_L^+, W_L^-, W_L^3$ | 1 | 3 | 1 | 0 |
| $W_R$ | $W_R^+, W_R^-, W_R^3$ | 1 | 1 | 3 | 0 |
| $B$ | $B$ | 1 | 1 | 1 | 0 |
| $\phi$ | $\begin{pmatrix} \phi_1^0 & \phi_1^+ \\ \phi_2^- & \phi_2^0 \end{pmatrix}$ | 1 | 2 | 2 | 0 |
| $\Delta_L$ | $\begin{pmatrix} \frac{\delta_L^+}{\sqrt{2}} & \delta_L^{++} \\ \delta_L^0 & -\frac{\delta_L^+}{\sqrt{2}} \end{pmatrix}$ | 1 | 3 | 1 | 2 |
| $\Delta_R$ | $\begin{pmatrix} \frac{\delta_R^+}{\sqrt{2}} & \delta_R^{++} \\ \delta_R^0 & -\frac{\delta_R^+}{\sqrt{2}} \end{pmatrix}$ | 1 | 1 | 3 | 2 |

**Table 1.** The field content in the LRSM and their corresponding quantum numbers. The fermionic fields $L$ and $Q$ represent lepton and quark fields, respectively. The index $i = 1, 2, 3$ of the fermionic fields runs over the number of generations. The $'$ symbol of the fermionic fields denotes that these are gauge eigenstates. The Higgs multiplets $\phi$, $\Delta_L$ and $\Delta_R$ are the bidoublet, the left and the right triplet, respectively.

While the VEVs $k_1$ and $k_2$ generate the SM gauge bosons $W_L$ and $Z_1$, the VEV $v_R$ gives rise to the new heavier and yet undiscovered gauge bosons $W_R$ and $Z_2$. The fact that heavier gauge bosons are yet undiscovered implies that

$$v_R \gg k_1, k_2. \tag{2.6}$$

Fermion masses are formed after SSB from the Yukawa interactions part of the LRSM Lagrangian. This part consists of the most general possible couplings of the Higgs multiplets to bilinear fermion field products which form singlets under $\mathrm{SU}(2)_L \times \mathrm{SU}(2)_R \times \mathrm{U}(1)_{B-L}$:

$$\mathcal{L}_Y = -\sum_{i,j}\Big[\bar{L}_{iL}\left((h_L)_{ij}\phi + (\tilde{h}_L)_{ij}\tilde{\phi}\right)L_{jR} - \bar{Q}_{iL}\left((h_Q)_{ij}\phi + (\tilde{h}_Q)_{ij}\tilde{\phi}\right)Q_{jR} \\ - \overline{(L_{iR})^c}\,\Sigma_R(h_M)_{ij}L_{jR} - \overline{(L_{iL})^c}\,\Sigma_L(h_M)_{ij}L_{jL}\Big] + h.c. \tag{2.7}$$

where $\tilde{\phi} \equiv \sigma_2\phi^*\sigma_2$, $\Sigma_{L,R} = i\sigma_2\Delta_{L,R}$ and $h_Q$, $h_L$,$h_M$,$\tilde{h_Q}$, $\tilde{h_L}$ are $3\times 3$ Yukawa matrices in flavor space (which have to be hermitian due to the left-right symmetry).[2] The gauge boson masses arise from the Higgs term in the Lagrangian, which consists of the Higgs kinetic terms and the potential of the Higgs multiplets:

$$\mathcal{L}_{\text{Higgs}} = \underbrace{\sum_i Tr|D_\mu\Theta_i|^2}_{\text{kinetic terms}} - \underbrace{V_{\text{Higgs}}}_{\text{potential}}\,, \tag{2.8}$$

where $\Theta_i = \{\phi, \Delta_L, \Delta_R\}$. After SSB, the charged and neutral gauge boson masses are generated through the above Higgs kinetic terms and are given by

$$M_{W_L}^2 \simeq \frac{g^2}{4}(k_1^2 + k_2^2), \qquad M_{W_R}^2 \simeq \frac{g^2 v_R^2}{2},$$
$$M_{Z_1}^2 \simeq \frac{g^2\,(k_1^2 + k_2^2)}{4\cos^2\theta_w}, \qquad M_{Z_2}^2 \simeq \frac{v_R^2 g^2\cos^2\theta_w}{cos2\theta_w}, \tag{2.9}$$

where $g \equiv g_L = g_R$ denote the SU(2) gauge couplings and $\theta_w$ is the weak mixing angle, defined in the SM as $e = g_L\sin\theta_w$ ($e$ being the electromagnetic coupling).[3]

## 3 LRSM constraints

We proceed with briefly summarizing the constraints on the relevant LRSM parameter space. We start with the quark masses, which are formed after SSB and are given by the Yukawa mass terms (see eq. (2.7)):

$$M_U = \frac{1}{\sqrt{2}}\Big(h_Q k_1 + \tilde{h}_Q k_2\Big),$$
$$M_D = \frac{1}{\sqrt{2}}\Big(h_Q k_2 + \tilde{h}_Q k_1\Big). \tag{3.1}$$

The observation that there is a quark mass scale in which the top quark mass is much larger than the bottom quark mass implies, assuming the absence of fine tuning, that $k_1$

[2] The manifest/quasi-manifest LRSM is realized upon assuming the left-right symmetry together with the lack of explicit CP violation in the Higgs potential. These assumptions imply, up to a sign, an identity between the corresponding elements of the left and the right CKM matrices (quasi-manifest LRSM). In this work we take the right CKM to be fully identical to the left CKM (manifest LRSM).

[3] We neglect here a small mixing angle and use the chiral eigenstates $W_{L,R}$ instead of the mass eigenstates $W_{1,2}$, respectively. The rotation (mixing angle) between the former and the latter respective states, $\xi$, is bound from above due to the Schwarz inequality [35]:

$$|\xi| \simeq \frac{|k_1 k_2|}{v_R^2} < \frac{M_{W_L}^2}{M_{W_R}^2} \lesssim 10^{-3}\,. \tag{2.10}$$

This small mixing is, as stated above, neglected here.

and $k_2$ are not of the same order (and neither are $h_Q$ and $\tilde{h}_Q$). In this work we therefore use $k_1 \gg k_2$, and in particular the natural setting $k_2/k_1 \sim m_b/m_t$ (see also [36]).

Additional constraints exist on the masses of the RH gauge boson $W_R$ and the heavy bidoublet Higgs bosons, and originate from flavor changing neutral current (FCNC) effects: $K - \bar{K}^0$, $B_{d,s} - \bar{B}^0_{d,s}$, $\epsilon_K$ mixings, as well as $b \to s\gamma$. These mixings receive significant contributions from the above mentioned particles and provide lower bounds on their masses: $M_{W_R} \geq 2.9\,\text{TeV}$, and $H_1^0, A_1^0 \gtrsim 10\,\text{TeV}$ [37–47]. A more stringent constraint on the mass of $W_R$ can be derived from calculating the radiative Higgs decay process $H \to \gamma\gamma$ [48]. This constraint, which depends on the doubly charged Higgs mass (chosen to be 0.8 TeV in this work), provides a lower bound of $M_{W_R} \gtrsim 3.8\,\text{TeV}$.[4] Direct lower bounds on $W_R$ mass are comparable with the theoretical bounds [54–57], whereas direct lower bounds on (heavy) neutral and singly charged Higgs bosons vary between $\sim 150\,\text{GeV} - 1.6\,\text{TeV}$, depending on the search mode for the neutral or singly charged Higgs sectors [58].

The direct bounds on the RH and LH doubly charged Higgs boson masses were obtained by searching for pair-produced "left-handed" states $\delta_L^{\pm\pm}$ and "right-handed" states $\delta_R^{\pm\pm}$:[5] $pp \to \delta^{++}_{L,R}\delta^{--}_{L,R} \to l_1^{\pm} l_2^{\pm} l_3^{\mp} l_4^{\mp}$ [59]. The lower bound on the $\delta_L^{\pm\pm}$ mass ranges from 770 GeV to 870 GeV where $B(\delta_L^{\pm\pm} \to l^{\pm}l^{\pm}) = 100\%$ and $l = e^{\pm}, \mu^{\pm}$ is assumed. In the case where $B(\delta_L^{\pm\pm} \to e^{\pm}e^{\pm}) = B(\delta_L^{\pm\pm} \to \mu^{\pm}\mu^{\pm}) = B(\delta_L^{\pm\pm} \to \tau^{\pm}\tau^{\pm}) \sim 33\%$ the lower bound varies between $\sim 650\,\text{GeV} - 680\,\text{GeV}$.[6] The lower bound on the mass of $\delta_R^{\pm\pm}$ varies from 660 GeV to 760 GeV, where again $l = e^{\pm}, \mu^{\pm}$ and $B(\delta_R^{\pm\pm} \to l^{\pm}l^{\pm}) = 100\%$ is assumed. This bound is similarly relaxed in the scenario (considered here) in which $B(\delta_R^{\pm\pm} \to e^{\pm}e^{\pm}) = B(\delta_R^{\pm\pm} \to \mu^{\pm}\mu^{\pm}) = B(\delta_R^{\pm\pm} \to \tau^{\pm}\tau^{\pm}) \sim 33\%$, where it varies between $\sim 500\,\text{GeV} - 600\,\text{GeV}$.

One particular constraint applies on the upper limit of the mass ratio between $N$ and $W_R$. Theoretical considerations based on stability and perturbativity of the effective potential (see [48] and also [60] for earlier results) suggest that it is possible for the heavy neutrinos in the LRSM to be heavier than $W_R$ depending on the measure of perturbativity within the LRSM parameter space. A ratio of

$$\frac{M_N}{M_{W_R}} \lesssim 7.3 \tag{3.2}$$

can be allowed without ruining neither stability nor perturbativity.[7]

[4]Extensions of the minimal LRSM which ameliorate indirect constraints are not ruled out. There are already a number of simple changes to the manifest LRSM which relax these constraints and allow coexistence of LHC detectable $W_R$ alongside lower limits on at least part of the Higgs sector (e.g. by adding extra $SU(2)_R$ quark doublet or Higgs multiplets to the model, by using a higher dimensional operator, or by differentiating the left and right couplings and mixing matrices — see [49–53]). These adjustments may add new processes which lead to the LSD+2 signal. We won't consider these cases here.

[5]The terms "left-handed" and "right-handed" describe the chirality (left or right) of the weak isospin $T_3$ coupled to the doubly charged Higgs: $\delta_L^{\pm\pm}$ coupled to either $l_L^- l_L'^-$ or $l_R^+ l_R'^+$ ($(T_3)_L = \pm 1$), and $\delta_R^{\pm\pm}$ coupled to either $l_R^- l_R'^-$ or $l_L^+ l_L'^+$ ($(T_3)_R = \pm 1$).

[6]The LH $H_L^{\pm\pm}$ does not contribute to the LSD + 2 jets signal due to vanishing of relevant couplings — see text.

[7]The measure of perurtbativity is the ratio between the self-generated 1-loop and the corresponding tree-level parameter [48]:

$$\frac{\alpha_3^{(1)}}{\alpha_3} = \frac{3\alpha_3}{8\pi^2}. \tag{3.3}$$

Eq. (3.2) corresponds to a maximal 100% perurbativity of $\alpha_3$.

Some constraints are required by unitarity. The optical theorem suggests the following bounds on the Higgs potential parameters: $\rho_{1,2} < 2\pi$ and $\alpha_3, \rho_3 < 8\pi$. In addition, the $\alpha_3$ parameter should satisfy the above mentioned FCNC constraints on the masses of $W_R$ and the heavy bidoublet Higgs bosons. In terms of the particle mass terms, the constraint is

$$M^2_{H_1^0,A_1^0} \simeq \frac{1}{2}\alpha_3 v_R^2 \frac{k_1^2+k_2^2}{k_1^2-k_2^2} \gtrsim 100\,\text{TeV}^2. \tag{3.4}$$

This leads to an interplay between the $\alpha_3$ parameter and the Higgs right triplet VEV $v_R$:

$$\alpha_3 \gtrsim \frac{2(k_1^2-k_2^2)}{(k_1^2+k_2^2)v_R^2} \cdot 100\,\text{TeV}^2. \tag{3.5}$$

The lower bound on $\alpha_3$ is thus directly related to the lower bound on the squared heavy neutral Higgs masses, and inversely related to the squared mass of $W_R$. Choosing the Higgs bidoublet VEVs so that $k_2/k_1 \sim m_b/m_t$ (see discussion below eq. (3.1)) and setting $\alpha_3 = 4.8$ ($\sim 18\%$ perturbativity, see footnote 7) yields $M_{W_R} \gtrsim 3\,\text{TeV}$, which is in agreement with [37–47].[8] This setting is used in section 4 to explore the LSD + 2 jets signal at the LHC (see also A).

It is worthwhile to explore the allowed parameter space constituting the mass term of the RH doubly charged Higgs $\delta_R^{\pm\pm}$ (which contributes to the LSD + 2 jets signal, as shown below). Incorporating the above constraints into a contour plot of the $\delta_R^{\pm\pm}$ mass term, which is given by

$$M^2_{\delta_R^{\pm\pm}} = 2\rho_2 v_R^2 + \frac{1}{2}\alpha_3(k_1^2-k_2^2), \tag{3.6}$$

yields the result illustrated in figure 1. The plot in this figure is given in the $\alpha_3 - \log_{10}\rho_2$ plane, where we have shaded the regions of the parameter space which are excluded by FCNC constraints (i.e. $M_{H_1^0,A_1^0} \gtrsim 10\,\text{TeV}$) and by direct LHC searches (the lower bound on $M_{\delta_R^{\pm\pm}}$). For clarity, the perturbativity measure of $\alpha_3$ is denoted (see footnote 7).

Further constraints are related to the Yukawa matrix $h_M$ which governs the couplings of the doubly charged Higgs to leptons (see below). The constraints on its elements originate from different low energy processes, such as $\mu \to \bar{e}ee$, Bhabha scattering, extra coupling to $(g-2)_\mu$, muonium ($\mu^+e^-$) transformation to anti-muonium and $\mu \to e\gamma$ decay [66–69].

Finally, relevant constraints also arise from the neutrinoless double $\beta$ decay process ($0\nu\beta\beta$) [6, 70–76]. Its non-observation sets a higher bound on the ratio between the mixing elements of the heavy Majorana neutrinos ($N_e, N_\mu, N_\tau$) and their masses [77, 78]:

$$\sum_{N=N_e,N_\mu,N_\tau} \frac{|K_{LNe}|^2}{M_N} < 5\times 10^{-8}\,\text{GeV}^{-1}, \tag{3.7}$$

[8]For a smaller measure of perturbativity, e.g. $\alpha_3 \sim 1$, eq. (3.5) gives $W_R \gtrsim 6\,\text{TeV}$, which is at the borderline of the LHC reach. In addition, larger values of $\alpha_3$ result in the $W_R$ and Higgs sector masses hitting the Landau pole at an energy scale which is much lower than the GUT scale, and therefore in the lack of ability to keep the scalar content of the LRSM reachable at the LHC [61–65]. That said, it is not ruled out that the minimal LRSM may be further extended at the TeV-scale (or a higher intermediate scale), such that all the gauge, scalar and Yukawa couplings are perturbative up to the GUT scale. There are already a number of simple mechanisms which relax indirect bounds and allow coexistence of LHC detectable $W_R$ alongside lower limits on at least part of the Higgs sector (see footnote 4).

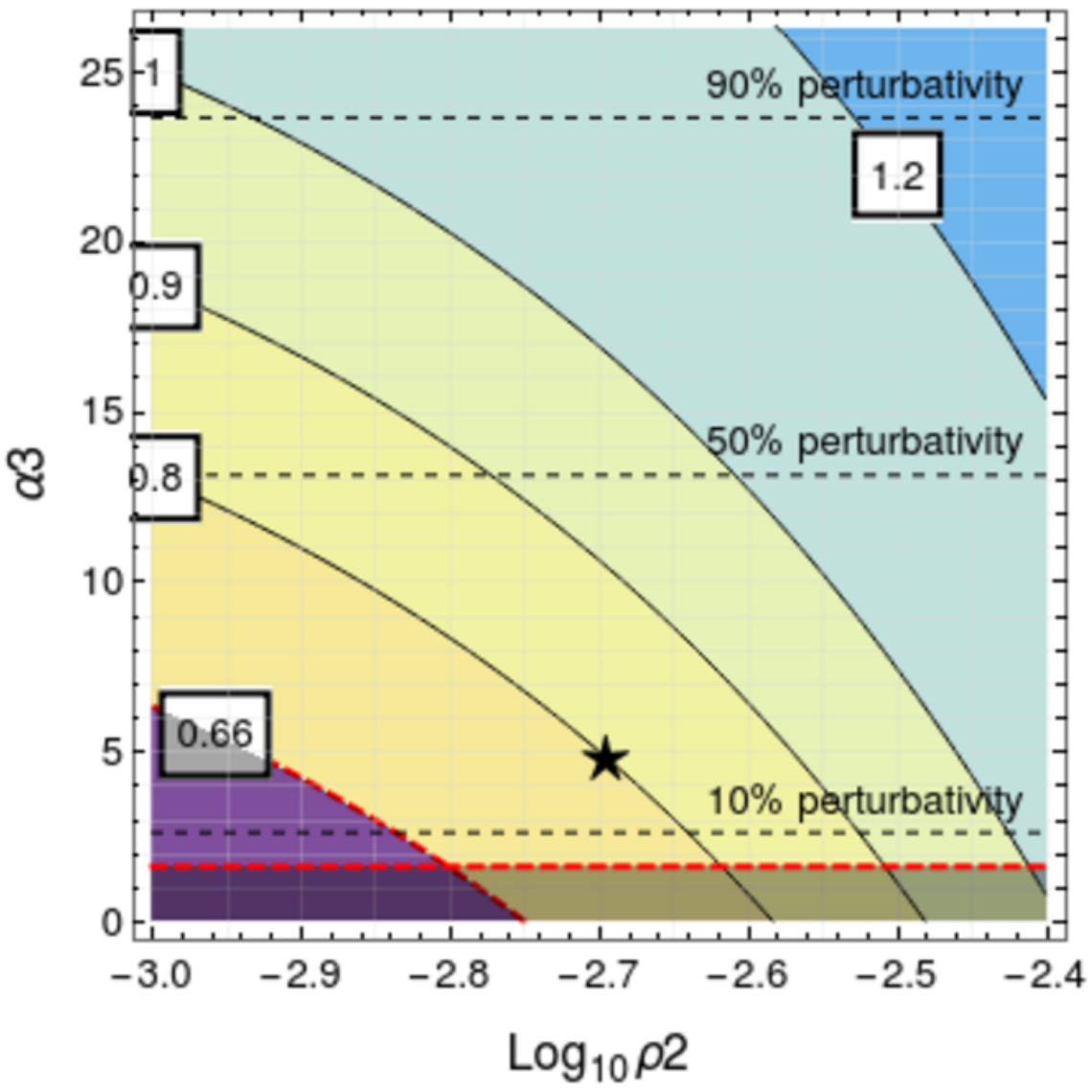

**Figure 1.** The mass of $\delta_R^{\pm\pm}$ (in TeV) as a function of the potential parameters $\alpha_3$ and $\log_{10}\rho_2$, with $M_{W_R} = 5.2\,\text{TeV}$. The contour plot is divided into mass regions shown in the legend. The shaded regions are excluded by FCNC constraints (the region under the lower horizontal dashed red line) and by direct LHC searches (the region left to the dashed red contour). The dashed black lines denote the $\alpha_3$ perturbativity percentage. The point marked by ★ corresponds to a parameter set used in the signal search of section 4 ($M_{W_R} = 5.2\,\text{TeV}$, $M_{\delta_R^{\pm\pm}} = 0.8\,\text{TeV}$, $\alpha_3 = 4.8$, $log_{10}\rho_2|_{M_{W_R}, M_{\delta_R^{\pm\pm}}, \alpha_3} = -2.695$ and $\alpha_3$ perturbativity of $\sim 18\%$).

where $K_L$ is a mixing matrix in the LH lepton charged current (see below).[9] Experimental limits are also extracted in present experiments by converting the sensitivity to the $0\nu\beta\beta$ process into a $0\nu\beta\beta$ decay strength parameter which has the dimension of mass. This parameter is denoted as the effective Majorana mass $m_{\beta\beta} \equiv |\sum_{\nu=\nu_e,\nu_\mu,\nu_\tau} K^2_{L\nu e} M_\nu|$ (where $\nu_e, \nu_\mu, \nu_\tau$ are the light Majorana neutrinos). Recent $0\nu\beta\beta$ searches involving detector materials such as Ge (GERDA, Majorana), Se (CUPID-0), Te (CUORE) and Xe (EXO-200, KamLAND-ZEN) [79–85] provide a combined range of upper limit bound on the effective Majorana mass: $m_{\beta\beta} < 0.1 - 0.23\,\text{eV}$, as well as upper limits on the lightest neutrino mass: $M_{\nu(\text{lightest})} < 0.15 - 0.44\,\text{eV}$ and on the sum of the light neutrino masses: $\Sigma \equiv \sum_{\nu=\nu_e,\nu_\mu,\nu_\tau} M_\nu < 0.46 - 1.3\,\text{eV}$.[10]

## 4 The LSD + 2 jets signal

The relevant $pp \to l^\pm l^\pm jj$ diagrams can be grouped according to two characteristics. The first is the kinematics structure of each diagram, which in this case can be either an s-channel or a t-channel. An additional distinctive feature is the presence of either a

[9] These constraints have been implemented in the manifest LRSM model file which will be later referred.

[10] The parameters of the manifest LRSM model file (see A) yield $m_{\beta\beta} \simeq 0.1\,\text{eV}$, $M_{\nu(\text{lightest})} = 0.1\,\text{eV}$ and $\Sigma = 0.3\,\text{eV}$, none of which is ruled out by the measurements.

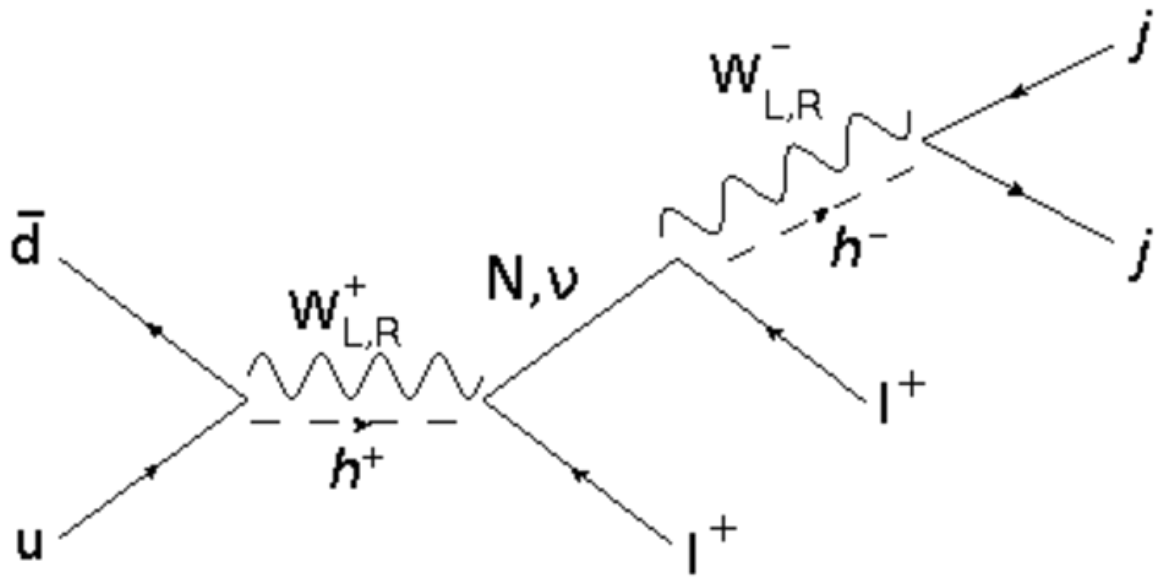

(a) Neutrino mediated processes.

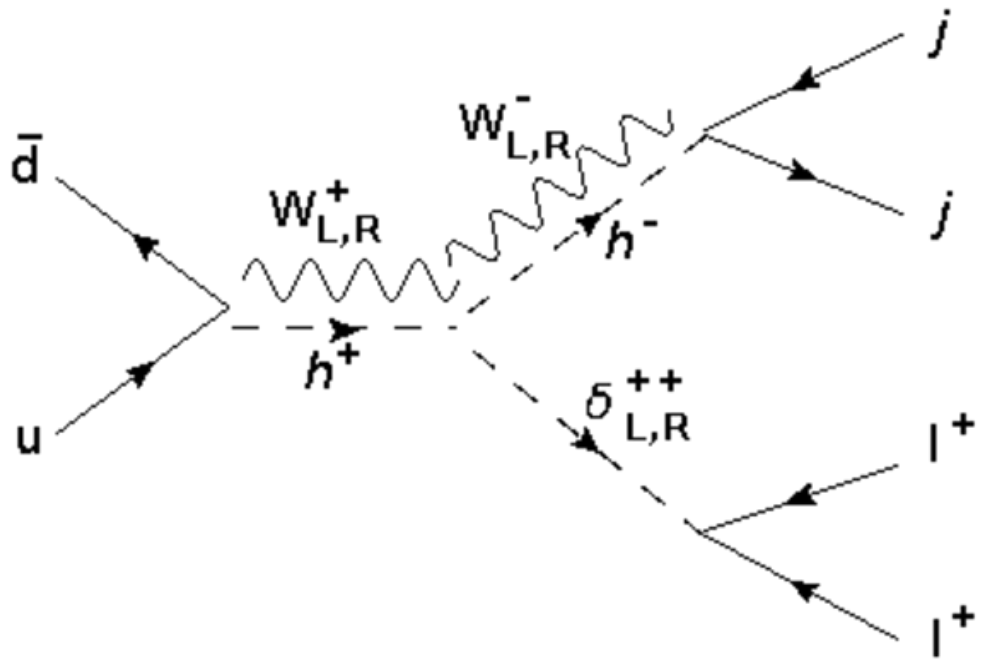

(b) Doubly charged Higgs mediated processes.

**Figure 2.** The s-channel diagrams leading to LSD + 2 jets signature. In each diagram, two lines with common endpoints represent two alternative possible particles, each with a different diagrammatic representation (names of particles with the same diagrammatic representation are separated by a comma). The generic $h^{\pm}$ field represents possible singly charged Higgs eigenstates.[13]

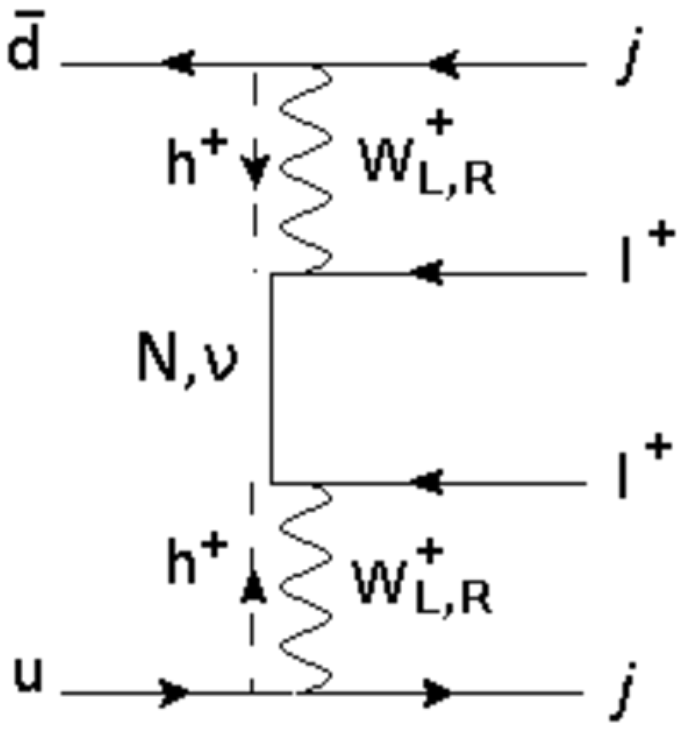

(a) Neutrino mediated processes.

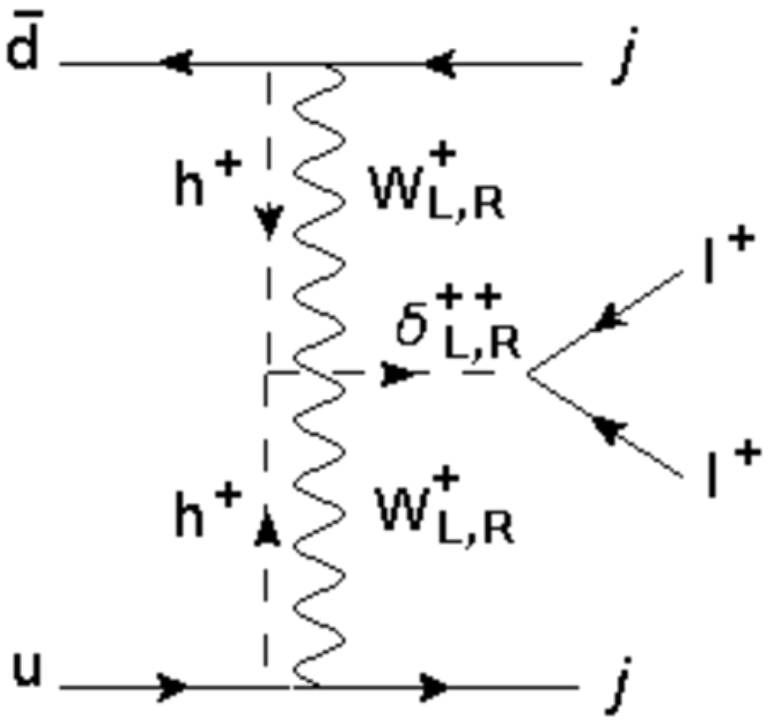

(b) Doubly charged Higgs mediated processes.

**Figure 3.** The t-channel processes leading to LSD + 2 jets signature. See also caption to figure 2.

mediating Majorana neutrino or a mediating doubly charged Higgs boson in the diagram.[11] This is illustrated in the subfigures of figure 2 (s-channel diagrams) and figure 3 (t-channel diagrams). We first note that for the tested parameters[12] the interference terms and t-channel diagrams were found to have a negligible contribution to the squared amplitude matrix of the $pp \to l^{\pm}l^{\pm}jj$ process. This contribution accumulates to a fraction of $\mathcal{O}(10^{-2})$ of the dominant, diagonal s-channel contributions. We therefore neglect it and focus on the latter. Within the s-channel group there are two dominant diagrams, one from each diagram type (neutrino mediated and doubly charged Higgs mediated), with significant contributions to the signal.

[11]We will therefore refer to each diagram as either neutrino mediated or doubly charged Higgs mediated.

[12]The relevant parameters are set subject to the constraints detailed in section 3 (see A). In addition, the RH neutrino masses are chosen to be degenerate and constrained to $M_N/M_{W_R} \leq 7.3$ (see text). The mass of the RH doubly charged Higgs $\delta_R^{++}$ is set to 800 GeV.

### 4.1 Doubly charged Higgs mediated diagrams

Almost all of the contributions from the s-channel doubly charged Higgs mediated diagrams (see figure 2b) can be safely neglected at the 14 TeV LHC with an integrated luminosity of 300 fb$^{-1}$. The only significant contribution from this group to the squared amplitude matrix belongs to the diagonal element of the (s-channel) process $pp \to W_R^+ \to W_R^{-*}\delta_R^{++*} \to e^+e^+jj$.[14] We first discuss its features, and later continue and shortly discuss the suppressing factors due to which the other processes shown in figure 2b were found to have negligible contributions to the signal.

**Significant contribution to the cross section: $\boldsymbol{pp \to W_R^+ \to W_R^{-*}\delta_R^{++*} \to e^+e^+jj}$.** The vertex factor of the $W_R$ production is governed by the covariant derivative in the fermion-gauge interaction term of the Lagrangian. Since the manifest LRSM consists of identical left and right CKM matrices as well as equal left and right gauge couplings ($g := g_L = g_R$), this vertex factor has identical size to its opposite parity SM analogue. The produced $W_R$ interacts via the $W_R^+W_R^+\delta_R^{--}$ vertex, which originates from the Higgs kinetic term $(D^\mu\Delta_R)^\dagger D_\mu\Delta_R$, and is given by the Lagrangian term

$$-\frac{1}{\sqrt{2}}g^2 v_R W_R^+ W_R^+ \delta_R^{--}. \tag{4.1}$$

The vertex factor (with $v_R$ of TeV scale) is significant. The $\delta_R^{++}$ propagates and decays into two positrons via an interaction which is governed by the Lagrangian term (see eq. (2.7) for the general Yukawa interaction)

$$\delta_R^{++}\overline{(e_R)^c}(h_M)_{ee}e_R, \tag{4.2}$$

where the Yukawa matrix $h_M$ is given by

$$h_M = \frac{1}{\sqrt{2}v_R}K_R^T M_{\text{diag}}^\nu K_R, \tag{4.3}$$

$K_R$ is a $6 \times 3$ mixing matrix for RH leptons[15] and $M_{\text{diag}}^\nu$ is the $6 \times 6$ neutrino diagonal mass matrix. We chose, as mentioned above, degenerate heavy neutrino masses (namely $M_N$). As a result we obtain for the positron-positron case

$$h_{Mee}(= h_{M\mu\mu} = h_{M\tau\tau}) \simeq 0.5\, g\, \frac{M_N}{M_{W_R}}. \tag{4.4}$$

This Yukawa coupling, in light of eq. (3.2), can be significant.

[13] While the manifest LRSM contains eight singly charged Higgs gauge eigenstates ($\delta_{L,R}^\pm, \phi_{1,2}^\pm$), in fact the only mass eigenstates which participate in the processes leading to the signal are $H_2^\pm$ (see B and discussion below).

[14] For the sake of simplicity and without loss of generality, we will deal in this subsection and the next one with diagrams leading to two positively charged leptons, although our discussion also applies to the sign-reversal diagrams. Moreover, although the expected cross section is similar for the three lepton generation pairs (as opposed to the background in which it is unnecessarily the same), we choose to deal with the first generation only in order to correspond with previous works related to this signal (see refs. [17, 18]).

[15] The block in $K_R$ connecting the heavy neutrino and charged leptons was chosen as identity block since the type I seesaw implies a heavy neutrino dominance in the RH charged current and a light neutrino dominance in the LH charged current. This setting is a simplification of the LH lepton matrix PMNS which has diagonal elements of $\mathcal{O}(1)$ — see [86]).

**Negligible contributions to the cross section: $pp \to W_R^+ \to H_2^{-*}\delta_R^{++*} \to e^+e^+jj$, $pp \to H_2^+ \to W_R^-\delta_R^{++} \to e^+e^+jj$.** These diagrams consist of the $W_R^+ H_2^+ \delta_R^{--}$ vertex[16] which is governed by the Higgs kinetic term $(D^\mu\Delta_R)^\dagger D_\mu \Delta_R$. The relevant Lagrangian term is

$$-ig\big[(\partial^\mu\delta_R^+)\delta_R^{--} - (\partial^\mu\delta_R^{--})\delta_R^+\big]\, W_R^+. \tag{4.5}$$

The electroweak coupling factor of the vertex does not counter-balance the negligibility of the diagrams, which results from the following: (i) the gauge eigenstate $\delta_R^\pm$ consists of a highly suppressed fraction of the physical (massive) Higgs eigenstate $H_2^\pm$:

$$\delta_R^\pm = \frac{1}{\sqrt{1+\frac{2(k_1^2+k_2^2)v_R^2}{(k_1^2-k_2^2)^2}}}\, H_2^\pm \tag{4.6}$$

(as mentioned above, natural setting is used: $k_2/k_1 \sim m_b/m_t$, and the coefficient goes as $\sim k_1/\sqrt{2}v_R$, $(< 1/30)$), (ii) the mass of $H_2^\pm$ is constrained by a lower limit. This is due to its dominant mass term (of $v_R$ scale) which is identical to the dominant term of the lower bounded heavy neutral Higgs particles $H_1^0$ and $A_1^0$ (see eq. (3.4)). Therefore, at the mass region in which the dominant channels can be discovered[17] — $H_2^\pm$ is significantly heavier than $W_R$ and its propagator factor is relatively suppressed, (iii) the coupling between $H_2^\pm$ and the proton quarks is proportional to the quarks masses, and thus very small (this applies only to the second process).

**$pp \to H_2^+ \to H_2^{-*}\delta_{L,R}^{++*} \to e^+e^+jj$.** In general, the triple-Higgs vertices involving one doubly charged Higgs and two singly charged Higgs bosons arise from the $\rho$, $\alpha$ and $\beta$ terms in the scalar potential:

$$\begin{aligned}
&\rho_1\left(\left(Tr\left[\Delta_L\Delta_L^\dagger\right]\right)^2+\left(Tr\left[\Delta_R\Delta_R^\dagger\right]\right)^2\right)+\rho_2\left(Tr[\Delta_L\Delta_L]Tr\left[\Delta_L^\dagger\Delta_L^\dagger\right]+Tr[\Delta_R\Delta_R]Tr\left[\Delta_R^\dagger\Delta_R^\dagger\right]\right)\\
&+\rho_3\left(Tr\left[\Delta_L\Delta_L^\dagger\right]Tr\left[\Delta_R\Delta_R^\dagger\right]\right)+\rho_4\left(Tr[\Delta_L\Delta_L]Tr\left[\Delta_R^\dagger\Delta_R^\dagger\right]+Tr\left[\Delta_L^\dagger\Delta_L^\dagger\right]Tr[\Delta_R\Delta_R]\right)\\
&+\alpha_1\left(Tr\left[\phi\phi^\dagger\right]\left(Tr\left[\Delta_L\Delta_L^\dagger\right]+\left[\Delta_R\Delta_R^\dagger\right]\right)\right)+\alpha_2\left(Tr\left[\phi\tilde{\phi}^\dagger\right]Tr\left[\Delta_R\Delta_R^\dagger\right]+Tr\left[\phi^\dagger\tilde{\phi}\right]Tr\left[\Delta_L\Delta_L^\dagger\right]\right)\\
&+\alpha_2^*\left(Tr\left[\phi^\dagger\tilde{\phi}\right]Tr\left[\Delta_R\Delta_R^\dagger\right]+Tr\left[\tilde{\phi}^\dagger\phi\right]Tr\left[\Delta_L\Delta_L^\dagger\right]\right)+\alpha_3\left(Tr\left[\phi\phi^\dagger\Delta_L\Delta_L^\dagger\right]+Tr\left[\phi^\dagger\phi\Delta_R\Delta_R^\dagger\right]\right)\\
&+\beta_1\left(Tr\left[\phi\Delta_R\phi^\dagger\Delta_L^\dagger\right]+Tr\left[\phi^\dagger\Delta_L\phi^\dagger\Delta_R^\dagger\right]\right)+\beta_2\left(Tr\left[\tilde{\phi}\Delta_R\phi^\dagger\Delta_L^\dagger\right]+Tr\left[\tilde{\phi}^\dagger\Delta_L\phi\Delta_R^\dagger\right]\right)\\
&+\beta_3\left(Tr\left[\phi\Delta_R\tilde{\phi}^\dagger\Delta_L^\dagger\right]+Tr\left[\phi^\dagger\Delta_L\tilde{\phi}\Delta_R^\dagger\right]\right).
\end{aligned} \tag{4.7}$$

These interaction terms are, at most, highly suppressed. We first note that the $\beta_i$ parameters of the manifest/quasi-manifest LRSM must vanish in order to reduce the mass scale of the non-SM gauge bosons without the need to fine-tune (and thus be able to theoretically allow possible observation at the LHC, see [9–11]). As for the $\rho_i$ and $\alpha_i$ terms in the potential, their contribution to the LSD+2 signal is negligible, the reason being is that all the relevant triple field products include the singly charged $\delta^{\pm}_{L,R}$ states. These states, in turn, comprise, respectively, the $H^{\pm}_1$ mass eigenstates, which are not part of any bidoublet eigenstate and therefore do not participate in the Yukawa coupling to the quarks, and a highly suppressed fraction of the heavy and tiny Yukawa coupled $H^{\pm}_2$ Higgs mass eigenstates (see discussion above and B).

[16]Replacing $W_R$ with the SM gauge boson interacting right-handedly results in a vertex with a small mixing angle factor. This factor, as explained in footnote 3, is ignored.

[17]The two dominant channels, i.e. the doubly charged Higgs mediated channel (discussed above) and the neutrino mediated channel (discussed below).

**$pp \to W^+_L \to W^{-*}_L \delta^{++*}_L \to e^+e^+jj$.** This diagram contains the LH doubly charged Higgs which interacts with two LH gauge bosons via the $W^+_L W^+_L \delta^{--}_L$ vertex. The relevant Lagrangian term is

$$-\frac{1}{\sqrt{2}} g^2 v_L W^+_L W^+_L \delta^{--}_L. \tag{4.8}$$

As this vertex factor is proportional to $v_L$, it vanishes in the framework of the manifest / quasi-manifest LRSM due to phenomenological considerations (see [9–11]).[18] In this case one cannot replace $W_L$ with a singly charged Higgs since, although a term $W^+_L \delta^+_L \delta^{--}_L$ appears in the Lagrangian, in the manifest LRSM the $\delta^{\pm}_L$ states are decoupled from the quarks (see above).

## 4.2 Majorana neutrino mediated diagrams

We now consider the (Majorana) neutrino mediated group of processes which lead to the LSD + 2 jets signal (see figure 2a). As discussed above, the processes we consider are characterized by an s-channel production of a (Majorana) neutrino and a positron/electron through an exchange of a gauge or a Higgs boson. The formed neutrino then decays into two jets plus a positron/electron — with an equal probability due to the Majorana nature of the neutrino (we are only concerned with signals which consist of two electrons or two positrons). As in the case of the doubly charged Higgs mediated diagrams, also many of the neutrino mediated diagrams can be safely neglected within the present LHC reach. We find that the significant contribution to the squared amplitude matrix arrives from the diagonal element which comprises the squared amplitude of the diagram $pp \to W^+_R \to e^+ N_e \to e^+e^+jj$ (as was first found by Keung and Senjanović, see [12]). We will shortly discuss this channel and the alternative suppressed channels in this diagram-group.

**Significant contribution to the cross section: $pp \to W^+_R \to e^+N_e \to e^+e^+jj$.** The produced $W_R$ decays into a positron[19] plus a heavy or a light electron-neutrino via the $\bar{N}_e W^+_R e$ or the $\bar{\nu}_e W^+_R e$ vertex, respectively (where $N_e$ ($\nu_e$) is the heavy (light) electron-neutrino element in a six-vector, $N$, constructed from three light and three heavy neutrino (Majorana) mass eigenstates). These vertices are governed by the lepton RH charged current terms:

$$L^e_{CC} = \frac{g}{\sqrt{2}} \left( \bar{N}_e \gamma^\mu K_{RN_e e}\, e_R W^+_{R\mu} + \bar{\nu}_e \gamma^\mu K_{R\nu_e e}\, e_R W^+_{R\mu} \right) + \text{h.c.} \tag{4.9}$$

[18] Replacing the SM $W_L$ with $W_R$ in the vertex will not help, as it leaves the $v_L$ untouched.

[19] The discussion in this subsection applies for sign reversal diagrams as well, see footnote 14.

(where $K_{L,R}$ are $6 \times 3$ left and right mixing matrices, respectively, in the lepton sector[20] (see also eq. (4.3))). The decay of $W_R$ through a RH current with a heavy neutrino is not suppressed for the chosen $K_R$ mixing matrix used here.[21] The probability of the alternative RH decay, through a light neutrino, is smaller by an order of $M_\nu/M_N$ (the quotient of the light and heavy neutrino masses).[22] The heavy neutrino (a Majorana particle) decays in equal rates into a same sign positron (plus two jets) and an opposite sign electron (plus two jets).[23]

**Negligible contributions to the cross section: $pp \to W_L^+ \to e^+(N,\nu)_e \to e^+e^+jj$.** The produced $W_L$ decays into a heavy or a light neutrino via the $\bar{N}_e W_L^+ e$ or $\bar{\nu}_e W_L^+ e$ vertex, respectively. The Lagrangian terms controlling these vertices are the chiral mirror of the terms related to the decay of $W_R$. They are given by

$$L_{CC}^e = \frac{g}{\sqrt{2}} \left( \bar{N}_e \gamma^\mu K_{L N_e e}\, e_L W_{L\mu}^+ + \bar{\nu}_e \gamma^\mu K_{L \nu_e e}\, e_L W_{L\mu}^+ \right) + \text{h.c.} \tag{4.10}$$

These two options are suppressed. The heavy neutrino as part of a LH charged current is suppressed due to the seesaw mechanism (see also footnote 22), and disfavored with a probability of $\mathcal{O}(M_\nu/M_N)$ in comparison to the light neutrino. But despite its dominance, the decay into the light neutrino also leads to a suppressed LSD production, the reason being is as follows. In terms of its helicity, a Majorana neutrino behaves at a weak current vertex as if it were a Dirac particle [87]. Thus, since the light neutrino is relativistic (but not massless), it is dominantly emitted alongside the positron at the $\bar{\nu}_e W_L^+ e$ vertex with a negative helicity, and has a highly suppressed probability (of $\mathcal{O}((M/E)^2)$) to be emitted in the "wrong" (positive) helicity. Now, as only the "wrong" helicity state can be absorbed without suppression at the next vertex ($\bar{\nu}_e W_L^+ e$, which emits a second positron at a LH charged current and therefore interacts significantly only for an incoming antineutrino with a positive helicity), this channel is therefore also highly suppressed.[24]

**$pp \to H_2^+ \to e^+(N,\nu)_e \to e^+e^+jj$.** The $H_2^+$ production and propagation parts of the diagram render it negligible. The $H_2^+$ production originate from the Yukawa couplings between the quarks and the bidoublet singly charged fields. These couplings are proportional to the masses of the light quarks within the proton (and are negligible in comparison to the weak coupling $g$ involved in the $W_{L,R}$ production). Moreover, the $H_2^+$ mass has a lower bound of approximately 10 TeV (see discussion above) — substantially heavier than an LHC detectable $W_R$.

[20]The mixing matrices $K_{L,R}$ contain both heavy-light mixing and mixing among generations. The inter-generation mixing elements in $K_R$ which connect heavy neutrinos and charged leptons (the non-diagonal elements in the block) are negelected here for the sake of simplicity — see also footnote 15.

[21]For instance, for the chosen parameter-set used here the number of produced heavy electron-neutrinos at the 14 TeV LHC for an integrated luminosity of $L = 300\,\text{fb}^{-1}$ is $\sim 2000$ for $(M_{W_R}, M_{M_N}) = (3\,\text{TeV}, 1\,\text{TeV})$ and $\sim 30$ for $(M_{W_R}, M_{M_N}) = (5.3\,\text{TeV}, 1\,\text{TeV})$.

[22]This is well demonstrated by the (type I) seesaw mechanism, where a diagonalization of a neutrino mass matrix composed of the masses related to the possible types of field billinear products gives eigenvalues of a light and a heavy neutrino masses, the geometric mean of which is the electroweak scale Dirac mass.

[23]There are two non-negligible decay branches of the heavy electron neutrino $N_e$ which lead to a second same sign lepton plus two jets: via RH and LH currents. While the LH current will be unfavored by a factor of $M_\nu/M_N$ because of the above mentioned seesaw mechanism, it is also enhanced due to the on-shell formation of $W_L$. Both of these decay branches are part of the $pp \to W_R^+ \to e^+N_e \to e^+e^+jj$ channel.

[24]While replacing the $W_L$ at the second vertex with $W_R$ eliminates this source of suppression, it in turn leads to a suppression due to the seesaw mechanism which favors a heavy neutrino in the RH current.

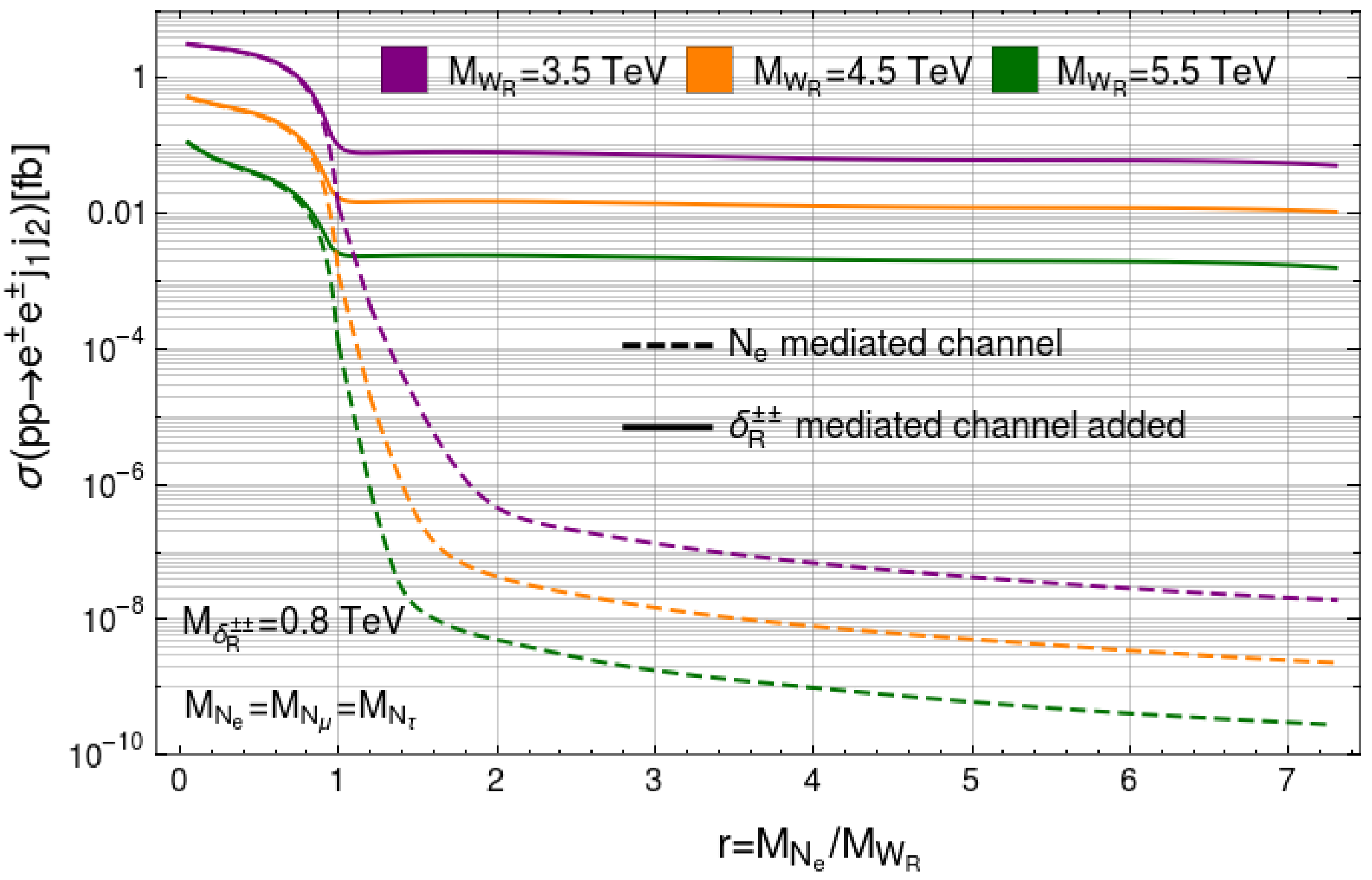

**Figure 4.** Cross sections for the $pp \to W_R^\pm \to e^\pm N_e \to e^\pm e^\pm jj$ processes (dashed lines) and the added contributions of the $pp \to W_R^\pm \to W_R^{\mp *} \delta_R^{\pm\pm *} \to e^\pm e^\pm jj$ processes (solid lines) at the 14 TeV LHC.

### 4.3 The cross section contributions from the two dominant diagrams

The cross-section arising from the two dominant diagonal terms of the squared amplitude matrix is plotted in figure 4. In the figure, and throughout this work, we have worked with degenerate heavy neutrinos and with positrons and electrons as $l^\pm l^\pm$. While the neutrino mediated channel is dominant at the lower $M_{N_e}/M_{W_R}$ range, it is inversely related to this ratio due to decreasing available phase space in the decay $W_R^\pm \to N_e\, e^\pm$ and due to the neutrino propagator factor which reduces the cross section for $M_{N_e} > M_{W_R}$. Conversely, the doubly charged Higgs mediated channel is enhanced as $M_{N_e}/M_{W_R}$ increases towards unity since the $W_R$ decay width decreases. As $M_{N_e}$ passes $M_{W_R}$ the decreasing of the $W_R$ decay width is stopped, and at that point the increasing decay width of $\delta_R^{++}$,[25] which reduces its propagator factor, starts to slowly reduce the channel cross-section as well.

[25] The Yukawa coupling of the $\delta_R^{++} ll$ vertex depends on $M_N/M_{W_R}$, as discussed above — see eq. (4.4).

| $(M_{W_R} = 4\,\text{TeV})$ | | $M_{N_e}/M_{W_R}$ | | |
|---|---|---|---|---|
| $(\delta_R^{\pm\pm} = 0.8\,\text{TeV})$ | 0.1 | 0.5 | 0.9 | 2.0 |
| $N_e$ channel | 6% | 42% | 43% | $--$ |
| $\delta_R^{\pm\pm}$ channel | 46% | 45% | 44% | 45% |

**Table 2.** The overall efficiency of the detector and the event selection criteria of the signal ($N_e$ and $\delta_R^{\pm\pm}$ mediated channels) for different values of $M_{N_e}/M_{W_R}$ ($M_{W_R} = 4\,\text{TeV}$, $M_{\delta_R^{\pm\pm}} = 0.8\,\text{TeV}$). The cross section for $M_{N_e}/M_{W_R} = 2$ in the $N_e$ mediated channel is too small for event generation.

### 4.4 Background analysis and sensitivity estimates

The SM processes which were considered as a potential background to the LSD + 2 jets signal are the ones leading to signatures with two electrons/positrons and at least two hadronic jets. The background processes which were considered are

$$pp \to \left\{ ZZ, ZW^{\pm}, t\bar{t} \to W^{\pm}W^{\mp}b\bar{b}, W^{\pm}W^{\pm}W^{\mp} \right\} \to e^{\pm}e^{\pm} + \text{jets} + \ldots \tag{4.11}$$

These processes do not violate lepton number and therefore contain (in addition to the signal ingredients) also opposite sign leptons and/or neutrinos in the final state. For generating signal and background events we used the `CalcHEP` [88] software with cteq6l1 parton distribution functions and an implementation of the manifest LRSM [89]. We used `PYTHIA` [90, 91] for the showering and hadronization routines. The K-factor for the signal was calculated using `FEWZ` (for the mass range of $M_{W_R} \sim 3-6$ TeV) [92].[26] For the detector simulation we used `DELPHES` [94] with ATLAS card (i.e. ATLAS detector specifications), and selected events with two isolated positrons/electrons and at least two jets in the final state. Processing the generated events was performed using the `MadGraph` [95] interface.

In order to reduce the background without considerably affecting the signal, the following cuts were applied on the selected events of the signal and the background:

- Each of the two jets with the highest[27] $E_T$ is required to have $E_T > 100\,\text{GeV}$,
- The invariant mass of the $ee$ system is required to be larger than $200\,\text{GeV}$,
- The $ee$ system is required to consist of either two positrons or two electrons.

The overall efficiency of the detector and the signal event selection is given in table 2 for the two dominant channels. The relatively low efficiency for $M_{N_e}/M_{W_R} = 0.1$ in the $N_e$ mediated channel is due to highly boosted $N_e$, leading to difficulties in separating its decay products in the detector. This feature does not occur in the $\delta_R^{\pm\pm}$ channel, where the kinematics of the signal particles is independent of $N_e$.

The signal and background events which survive the event selection and subsequently pass the above cuts are tagged as S and B, respectively, and are used as a common event pool. In order to reconstruct the relevant LRSM particles, the pool events are probed for the following invariant mass systems: $e_1^\pm e_2^\pm j_1 j_2$ (which corresponds to reconstructing the $W_R$ mass), $e^\pm j_1 j_2$ (the $N_e$ mass reconstruction) and $e_1^\pm e_2^\pm$ (the $\delta_R^{\pm\pm}$ mass reconstruction). The reconstructed LRSM masses associated with three chosen mass coordinates are shown in figure 5. In the figure we plot the number of events as a function of the reconstructed masses of $W_R^\pm$ (top panel), $N_e$[28] (middle panel) and $\delta_R^{\pm\pm}$ (bottom panel). It is evident from the plots that the background becomes low before the signal peak for each of the three particle masses, which come out clear and distinct within the reach of the LHC.

[26] The K-factor is 1.3 for the signal and 1.4 for the background processes (see also [93]).
[27] We assume that the two leading jets come from $N_e$.

In order to map the mass region with sensitivity to the signal we proceed as follows: in every mass coordinate tested for the signal we use the available detectable systems (i.e. $e_1^\pm e_1^\pm j_1 j_2$, $e^\pm j_1 j_2$ and $e_1^\pm e_2^\pm$) to compare the signal and the background events which pass the above event selection and cuts. We apply the two following selection rules on the available S and B counts for each system (in a selected mass window, see ref. [17]):

1. The presence of at least 10 signal events, and
2. The signal must exceed five statistical fluctuations of the background ($\frac{S}{\sqrt{B}} \geq 5$).

We require at least one of the above systems to pass both of these selection rules to allow discovery. For a given mass setting, passing the discovery criterion is related to the joint contribution of the two channels. In addition, we check whether the discovery can be attributed to any of the two channels independently. We continue and map the region boundaries of the independent and the joint (combined) channels with corresponding discovery contours in the ($M_{W_R}$, $M_{N_e}$) plane (with a chosen $M_{\delta_R^{\pm\pm}} = 0.8\,\text{TeV}$). The constructed discovery contours are shown in figure 6. They measure the detector sensitivity to the two channels, both separately and jointly (combined).

After data-taking at an integrated luminosity of $300\,\text{fb}^{-1}$, the discovery limits reach a maximal $M_{W_R}$ of 5.3 TeV and a maximal $M_{N_e}$ of 3.6 TeV for the $N_e$ mediated channel alone.

Upon adding the contribution of the $\delta_R^{\pm\pm}$ mediated channel, the maximal $M_{W_R}$ limit associated with the $N_e$ mediated channel is pushed out marginally by 30 GeV at $M_{N_e} \simeq 1.6\,\text{TeV}$. For $N_e$ masses below 1.6 TeV the $M_{N_e}$ discovery limit is pushed out by less than 30 GeV. For $N_e$ masses above 1.6 TeV the upper $M_{N_e}$ discovery limit[29] for $W_R$ masses going down to 4.4 TeV rises by up to $\sim 100\,\text{GeV}$. As the mass of $W_R$ gets nearer (and then below) 4 TeV the significance of the $\delta_R^{\pm\pm}$ channel markedly grows[30] and the upper $N_e$ mass discovery limit starts increasing rapidly as $M_{W_R}$ declines. This rapid increase reaches an upper discovery limit of $M_{N_e} \sim 33\,\text{TeV}$ for $M_{W_R} = 3\,\text{TeV}$, beyond the allowed theoretical limit (see eq. (3.2)).

As a more general (and clarifying) perspective to the above discussion it is beneficial to demonstrate the ultimate contribution of the Higgs mediated channel to the signal in terms of the number of detected events as a function of $W_R$ and $\delta_R^{\pm\pm}$ masses in a two-dimensional $(M_{W_R}, M_{\delta_R^{\pm\pm}})$ mass plane area. This is shown in figure 7, for an integrated luminosity of $300\,\text{fb}^{-1}$. As discussed in subsection 4.3, the maximal cross section in each $(M_{W_R}, M_{\delta_R^{\pm\pm}})$ coordinate occurs for $M_{N_e} = M_{W_R}$ (yet it decreases slowly and retains much of its relative strength throughout considerably heavier $N_e$ mass range, as shown in figure 4.).[31] The figure demonstrates, particularly for the lower mass areas (and at this chosen luminosity), that the Higgs mediated channel can be discovered either independently or jointly with the neutrino mediated channel.

---

[28] Since we were unable to deductively identify the lepton which originates from the $N_e$ decay, both combinations of $e^\pm j_1 j_2$ had to be examined (resulting in a wider mass spectra).

[29] As shown in figure 6, for each given $W_R$ mass there is a lower and a higher $N_e$ mass discovery limit.

[30] The number of events from the $N_e$ mediated channel becomes negligible as $M_{N_e}$ approaches and passes $M_{W_R}$ for a 14 TeV LHC and an integrated luminosity of $300\,\text{fb}^{-1}$.

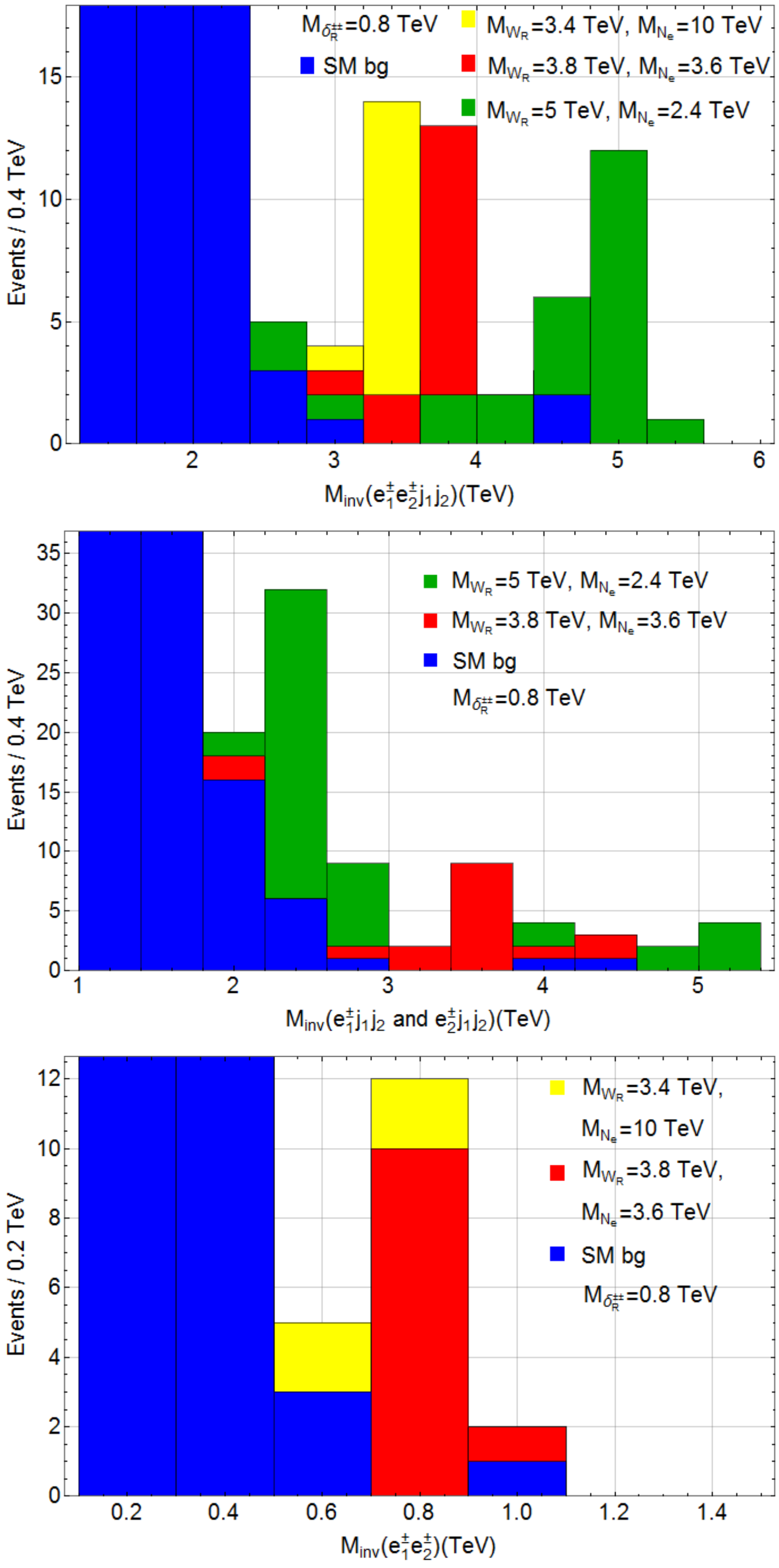

**Figure 5.** Mass reconstructions of $M_{W_R}$, $M_{N_e}$ and $M_{\delta_R^{\pm\pm}}$, together with SM background at the 14 TeV LHC. The reconstructions are performed with the signal and background selected events as a function of the invariant mass of the $e_1^\pm e_1^\pm j_1 j_2$ system (top panel), the $e_1^\pm j_1 j_2$ and $e_2^\pm j_1 j_2$ systems (middle panel) and the $e_1^\pm e_2^\pm$ system (bottom panel), for an integrated luminosity of 300 fb$^{-1}$. The three mass sets used in this figure are marked in figure 6.

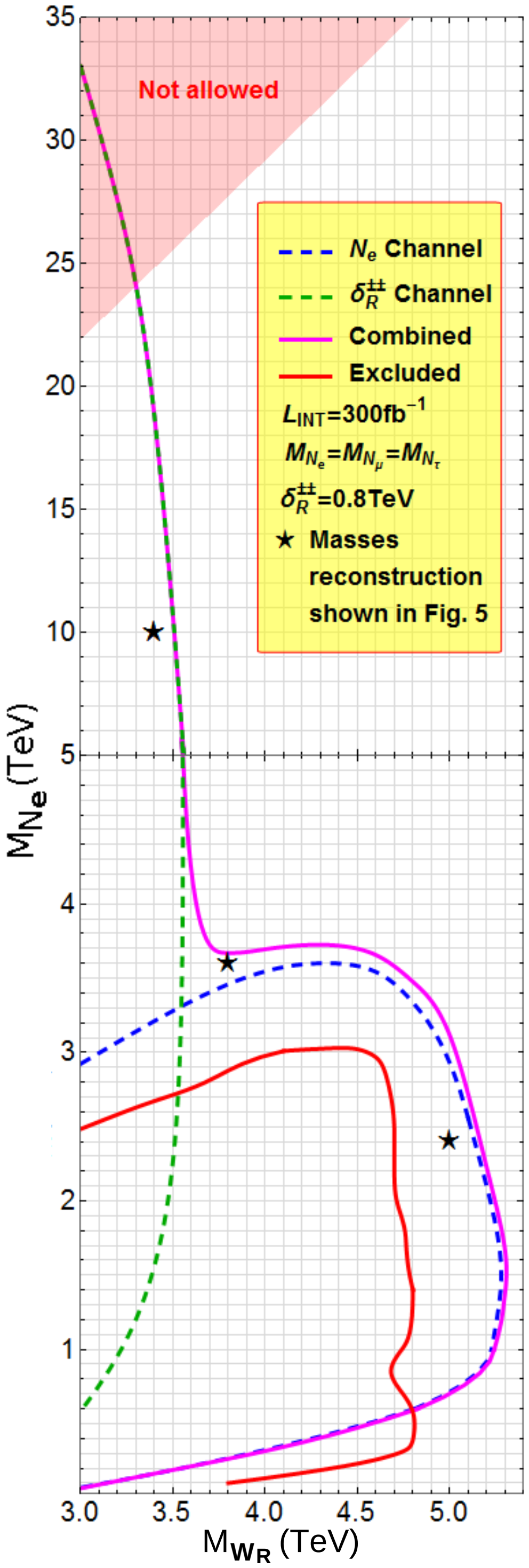

**Figure 6.** Discovery potential for the two dominant channels of the $pp \to W_R \to e^\pm e^\pm jj$ process at the 14 TeV LHC for an integrated luminosity of 300 fb$^{-1}$. The coordinates denoted by ★ correspond to the three mass settings of figure 5. The red line boundary of the excluded area (i.e. the mass region where no excess of like-sign dilepton events in comparison to the SM was found) is based on the works in refs. [54–57]. The lower and the upper part of the y-axis have different (linear) scales in order to highlight the independent and the combined sensitivity regions associated with the two channels.

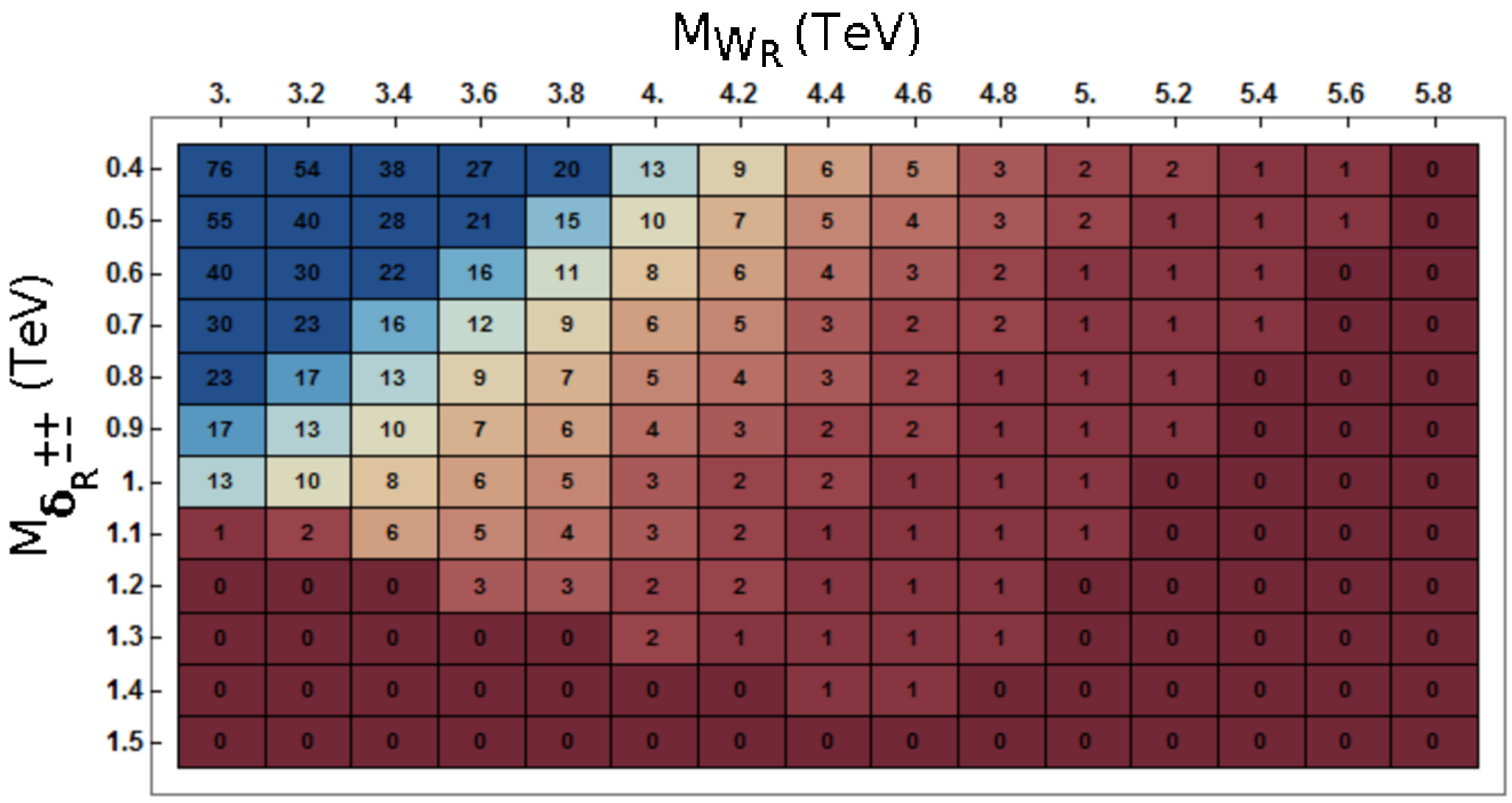

**Figure 7.** Maximal number of detected events by the Higgs mediated channel for $(M_{W_R}, M_{\delta_R^{\pm\pm}})$ mass coordinates in a two-dimensional mass plane area (for an integrated luminosity of $300\,\text{fb}^{-1}$. The efficiency is given in table 2).

## 5 Summary

The importance of the LRSM lies mainly in the fact that it restores parity symmetry at higher energy scales and provides a natural setup for the observed neutrino oscillations phenomena, based on the see-saw mechanism. One particular process within the framework of the LRSM which is able to supply direct evidence to both the left-right symmetry breaking scale and the correctness of the see-saw mechanism is $pp \to e^{\pm}e^{\pm}jj + X$, generating a signal of two same sign leptons (electrons/positrons in this work) plus two jets. We evaluated the cross sections of the dominant process channels leading to this signal at the 14 TeV LHC, with an integrated luminosity of $300\,\text{fb}^{-1}$ — for a chosen set of parameters. We estimated the mass regions in which the LHC is sensitive to the two dominant process channels, and showed that the contribution of the doubly charged Higgs mediated channel can significantly expand the discovery potential which arises from the neutrino mediated channel alone.

[31] We remind that the masses of the three heavy neutrinos are restricted to be equal in this work.

## A Parameter settings used

The manifest LRSM model file used in this work is described in ref. [89], and is based on the model as described in ref. [11]. The parameter settings used are as follows:

- Higgs VEVs (in GeV)

$$k_1 = 246, \qquad k_2 = \sqrt{246.22^2 - k_1^2} = 10.41,$$
$$v_R = 6386 \ldots 12772.$$

- Parameters in the Higgs potential

$$\lambda_1 = 0.1, \quad \lambda_2 = 0.025, \quad \lambda_3 = 0.05, \quad \lambda_4 = 0.05,$$
$$\rho_1 = 0.9, \quad \rho_3 = 1.81, \quad \rho_4 = 1,$$
$$\rho_2 = 0.00151 \ldots 0.00606,$$
$$\alpha_1 = 0.5, \quad \alpha_2 = 0.5, \quad \alpha_3 = 4.8.$$

- Couplings

$$G_f = 1.166 \times 10^{-5}\,\mathrm{GeV}^{-2}, \qquad \alpha_s(M_Z) = 0.1184, \qquad \alpha(0) = \frac{1}{137.036},$$

- Light neutrino masses: $M_{\nu_e} = M_{\nu_\mu} = M_{\nu_\tau} = 0.1\,\mathrm{ev}$,
  Heavy neutrino masses: $M_{N_e} = M_{N_\mu} = M_{N_\tau}$,

- Lepton mixing matrices: $K_{L_{i,j}}$, $K_{R_{i,j}}$[32] $(i = 1 \ldots 6,\ j = 1 \ldots 3)$.

$$K_{Li,j} = PMNS_{i,j}\ (i = 1 \ldots 3),$$[33]

$$K_{L4,1} = V_e, \quad K_{L5,2} = V_\mu, \quad K_{L6,3} = V_\tau, \quad K_{Li \neq j+3} = 0\ (i = 4 \ldots 6),$$
$$K_{R1,1} = -V_e, \quad K_{R2,2} = -V_\mu, \quad K_{R3,3} = -V_\tau, \quad K_{Ri \neq j} = 0\ (i = 1 \ldots 3),$$
$$K_{Ri=j+3} = 1, \quad K_{Ri \neq j+3} = 0\ (i = 4 \ldots 6),$$
$$V_e = \sqrt{M_{\nu_e}/M_{N_e}}, \quad V_\mu = \sqrt{M_{\nu_\mu}/M_{N_\mu}}, \quad V_\tau = \sqrt{M_{\nu_\tau}/M_{N_\tau}}.$$

## B Higgs physical eigenstates

The Higgs multiplets consist of 20 degrees of freedom, i.e. 20 real fields. Obtaining the fields eigensystem is done by diagonalizing the squared-mass matrix:

$$\frac{\partial^2}{\partial\phi_i \partial\phi_j} V \Big|_{\phi_i = \phi_j = 0} = m_{i,j}^2. \tag{B.1}$$

The eigenstates consist of[34]

[32]The lepton mixing parameters $K_{L,R}$ are $6 \times 3$ matrices in the lepton sector which connect the charged leptons to the six Majorana neutrinos.
[33]The PMNS matrix is the lepton sector analogue to the CKM quark matrix (see [86]).
[34]We use the unitary gauge.

1. Four neutral scalar eigenstates $H$, $H_1^0$, $H_2^0$, $H_3^0$,
2. Four neutral pseudoscalar eigenstates $A_1^0$, $A_2^0$, $G_1^0$, $G_2^0$ ($G_1^0$ and $G_2^0$ are Goldstone bosons),
3. Four singly charged scalar eigenstates $H_1^\pm$, $H_2^\pm$, $G_L^\pm$, $G_R^\pm$ ($G_L^\pm$ and $G_R^\pm$ are Goldstone bosons),
4. Two doubly charged scalar eigenstates $H_L^{\pm\pm}$, $H_R^{\pm\pm}$.

The corresponding eigenvalues/masses are given, e.g., in the third ref. in [9–11]. The non-physical Higgs fields may be written in terms of the above eigenstates as follows (the $\phi_1^0$, $\phi_2^0$ and $\delta_R^0$ states are given in the approximation $v_R \gg k_+$, where $k_\pm \equiv \sqrt{k_1^2 \pm k_2^2}$):

$$
\begin{aligned}
\phi_1^0 &\approx \frac{1}{k_+\sqrt{2}}\left(k_1 k_+ + k_1 H - k_2 H_1^0 - i k_1 G_1^0 + i k_2 A_1^0\right), \\
\phi_2^0 &\approx \frac{1}{k_+\sqrt{2}}\left(k_2 k_+ + k_2 H + k_1 H_1^0 + i k_2\, G_1^0 + i k_1\, A_1^0\right), \\
\delta_L^0 &= \frac{1}{\sqrt{2}}\left(v_L + H_3^0 + i\, A_2^0\right), \\
\delta_R^0 &\approx \frac{1}{\sqrt{2}}\left(v_R + H_2^0 + i\, G_2^0\right), \\
\phi_1^\pm &= \frac{k_1}{k_+\sqrt{1+(\frac{k_-^2}{\sqrt{2}k_+ v_R})^2}}\, H_2^\pm - \frac{k_1}{k_+\sqrt{1+(\frac{\sqrt{2}k_+ v_R}{k_-^2})^2}}\, G_R^\pm - \frac{k_2}{k_+}\, G_L^\pm \\
\phi_2^\pm &= \frac{k_2}{k_+\sqrt{1+(\frac{k_-^2}{\sqrt{2}k_+ v_R})^2}}\, H_2^\pm - \frac{k_2}{k_+\sqrt{1+(\frac{\sqrt{2}k_+ v_R}{k_-^2})^2}}\, G_R^\pm + \frac{k_1}{k_+}\, G_L^\pm \\
\delta_L^\pm &= H_1^\pm, \\
\delta_R^\pm &= \frac{1}{\sqrt{1+(\frac{\sqrt{2}k_+ v_R}{k_-^2})^2}}\, H_2^\pm + \frac{1}{\sqrt{1+(\frac{k_-^2}{\sqrt{2}k_+ v_R})^2}}\, G_R^\pm \\
\delta_{L,R}^{\pm\pm} &= H_{L,R}^{\pm\pm}.
\end{aligned}
\tag{B.2}
$$

## References

[1] J.C. Pati and A. Salam, *Lepton number as the fourth color*, *Phys. Rev. D* **10** (1974) 275 [*Erratum ibid.* **11** (1975) 703] [INSPIRE].

[2] R.N. Mohapatra and J.C. Pati, *Left-right gauge symmetry and an isoconjugate model of CP-violation*, *Phys. Rev. D* **11** (1975) 566 [INSPIRE].

[3] G. Senjanović and R.N. Mohapatra, *Exact left-right symmetry and spontaneous violation of parity*, *Phys. Rev. D* **12** (1975) 1502 [INSPIRE].

[4] P. Minkowski, *$\mu \to e\gamma$ at a rate of one out of $10^9$ muon decays?*, Phys. Lett. B **67** (1977) 421 [INSPIRE].

[5] R.N. Mohapatra and G. Senjanović, *Neutrino mass and spontaneous parity nonconservation*, Phys. Rev. Lett. **44** (1980) 912 [INSPIRE].

[6] R.N. Mohapatra and G. Senjanović, *Neutrino masses and mixings in gauge models with spontaneous parity violation*, Phys. Rev. D **23** (1981) 165 [INSPIRE].

[7] G. Senjanović, *Seesaw at LHC through left-right symmetry*, Int. J. Mod. Phys. A **26** (2011) 1469 [arXiv:1012.4104] [INSPIRE].

[8] Super-Kamiokande collaboration, *Evidence for oscillation of atmospheric neutrinos*, Phys. Rev. Lett. **81** (1998) 1562 [hep-ex/9807003] [INSPIRE].

[9] J.F. Gunion, J. Grifols, A. Mendez, B. Kayser and F.I. Olness, *Higgs bosons in left-right symmetric models*, Phys. Rev. D **40** (1989) 1546 [INSPIRE].

[10] N.G. Deshpande, J.F. Gunion, B. Kayser and F.I. Olness, *Left-right symmetric electroweak models with triplet Higgs*, Phys. Rev. D **44** (1991) 837 [INSPIRE].

[11] P. Duka, J. Gluza and M. Zralek, *Quantization and renormalization of the manifest left-right symmetric model of electroweak interactions*, Annals Phys. **280** (2000) 336 [hep-ph/9910279] [INSPIRE].

[12] W.-Y. Keung and G. Senjanović, *Majorana neutrinos and the production of the right-handed charged gauge boson*, Phys. Rev. Lett. **50** (1983) 1427 [INSPIRE].

[13] W.H. Furry, *On transition probabilities in double beta-disintegration*, Phys. Rev. **56** (1939) 1184 [INSPIRE].

[14] C.E. Picciotto and M.S. Zahir, *Neutrinoless double beta decay in left-right symmetric models*, Phys. Rev. D **26** (1982) 2320 [INSPIRE].

[15] K. Huitu, J. Maalampi, A. Pietila and M. Raidal, *Doubly charged Higgs at LHC*, Nucl. Phys. B **487** (1997) 27 [hep-ph/9606311] [INSPIRE].

[16] J. Maalampi and N. Romanenko, *Single production of doubly charged Higgs bosons at hadron colliders*, Phys. Lett. B **532** (2002) 202 [hep-ph/0201196] [INSPIRE].

[17] A. Ferrari et al., *Sensitivity study for new gauge bosons and right-handed Majorana neutrinos in pp collisions at $\sqrt{s} = 14$ TeV*, Phys. Rev. D **62** (2000) 013001 [INSPIRE].

[18] S.N. Gninenko, M.M. Kirsanov, N.V. Krasnikov and V.A. Matveev, *Detection of heavy Majorana neutrinos and right-handed bosons*, Phys. Atom. Nucl. **70** (2007) 441 [INSPIRE].

[19] A. Atre, T. Han, S. Pascoli and B. Zhang, *The search for heavy Majorana neutrinos*, JHEP **05** (2009) 030 [arXiv:0901.3589] [INSPIRE].

[20] M. Nemevšek, F. Nesti, G. Senjanović and Y. Zhang, *First limits on left-right symmetry scale from LHC data*, Phys. Rev. D **83** (2011) 115014 [arXiv:1103.1627] [INSPIRE].

[21] C.-Y. Chen and P.S. Dev, *Multi-lepton collider signatures of heavy Dirac and Majorana neutrinos*, Phys. Rev. D **85** (2012) 093018 [arXiv:1112.6419] [INSPIRE].

[22] J. Chakrabortty, J. Gluza, R. Sevillano and R. Szafron, *Left-right symmetry at LHC and precise 1-loop low energy data*, JHEP **07** (2012) 038 [arXiv:1204.0736] [INSPIRE].

[23] J.A. Aguilar-Saavedra and F.R. Joaquim, *Measuring heavy neutrino couplings at the LHC*, Phys. Rev. D **86** (2012) 073005 [arXiv:1207.4193] [INSPIRE].

[24] T. Han, I. Lewis, R. Ruiz and Z.-G. Si, *Lepton number violation and $W'$ chiral couplings at the LHC*, *Phys. Rev. D* **87** (2013) 035011 [*Erratum ibid.* **87** (2013) 039906] [arXiv:1211.6447] [INSPIRE].

[25] C.-Y. Chen, P.S.B. Dev and R.N. Mohapatra, *Probing heavy-light neutrino mixing in left-right seesaw models at the LHC*, *Phys. Rev. D* **88** (2013) 033014 [arXiv:1306.2342] [INSPIRE].

[26] T.G. Rizzo, *Exploring new gauge bosons at a* 100 *TeV collider*, *Phys. Rev. D* **89** (2014) 095022 [arXiv:1403.5465] [INSPIRE].

[27] J. Gluza and T. Jeliński, *Heavy neutrinos and the $pp \to lljj$ CMS data*, *Phys. Lett. B* **748** (2015) 125 [arXiv:1504.05568] [INSPIRE].

[28] J.N. Ng, A. de la Puente and B.W.-P. Pan, *Search for heavy right-handed neutrinos at the LHC and beyond in the same-sign same-flavor leptons final state*, *JHEP* **12** (2015) 172 [arXiv:1505.01934] [INSPIRE].

[29] P.S.B. Dev, D. Kim and R.N. Mohapatra, *Disambiguating seesaw models using invariant mass variables at hadron colliders*, *JHEP* **01** (2016) 118 [arXiv:1510.04328] [INSPIRE].

[30] M. Mitra, R. Ruiz, D.J. Scott and M. Spannowsky, *Neutrino jets from high-mass $W_R$ gauge bosons in TeV-scale left-right symmetric models*, *Phys. Rev. D* **94** (2016) 095016 [arXiv:1607.03504] [INSPIRE].

[31] A. Das, P.S.B. Dev and R.N. Mohapatra, *Same sign versus opposite sign dileptons as a probe of low scale seesaw mechanisms*, *Phys. Rev. D* **97** (2018) 015018 [arXiv:1709.06553] [INSPIRE].

[32] C. Arbelaéz, C. Dib, I. Schmidt and J.C. Vasquez, *Probing the Dirac or Majorana nature of the heavy neutrinos in pure leptonic decays at the LHC*, *Phys. Rev. D* **97** (2018) 055011 [arXiv:1712.08704] [INSPIRE].

[33] M. Nemevšek, F. Nesti and G. Popara, *Keung-Senjanović process at the LHC: from lepton number violation to displaced vertices to invisible decays*, *Phys. Rev. D* **97** (2018) 115018 [arXiv:1801.05813] [INSPIRE].

[34] A. Davidson, *$B - L$ as the fourth color within an $\mathrm{SU}(2)_L \times \mathrm{U}(1)_R \times \mathrm{U}(1)$ model*, *Phys. Rev. D* **20** (1979) 776 [INSPIRE].

[35] J. Chay, K.Y. Lee and S.-H. Nam, *Bounds of the mass of $Z'$ and the neutral mixing angles in general $\mathrm{SU}(2)_L \times \mathrm{SU}(2)_R \times \mathrm{U}(1)$ models*, *Phys. Rev. D* **61** (2000) 035002 [hep-ph/9809298] [INSPIRE].

[36] Y. Zhang, H. An, X. Ji and R.N. Mohapatra, *General CP-violation in minimal left-right symmetric model and constraints on the right-handed scale*, *Nucl. Phys. B* **802** (2008) 247 [arXiv:0712.4218] [INSPIRE].

[37] G. Beall, M. Bander and A. Soni, *Constraint on the mass scale of a left-right symmetric electroweak theory from the $K_L - K_S$ mass difference*, *Phys. Rev. Lett.* **48** (1982) 848 [INSPIRE].

[38] G. Ecker, W. Grimus and H. Neufeld, *Higgs induced flavor changing neutral interactions in* $\mathrm{SU}(2)_L \times \mathrm{SU}(2)_R \times \mathrm{U}(1)$, *Phys. Lett. B* **127** (1983) 365 [*Erratum ibid.* **132** (1983) 467] [INSPIRE].

[39] A. Maiezza, M. Nemevšek, F. Nesti and G. Senjanović, *Left-right symmetry at LHC*, *Phys. Rev. D* **82** (2010) 055022 [arXiv:1005.5160] [INSPIRE].

[40] M. Blanke, A.J. Buras, K. Gemmler and T. Heidsieck, $\Delta F = 2$ *observables and* $B \to X_q\gamma$ *decays in the left-right model: Higgs particles striking back*, JHEP **03** (2012) 024 [arXiv:1111.5014] [INSPIRE].

[41] S. Bertolini, A. Maiezza and F. Nesti, *Present and future* $K$ *and* $B$ *meson mixing constraints on TeV scale left-right symmetry*, Phys. Rev. D **89** (2014) 095028 [arXiv:1403.7112] [INSPIRE].

[42] V. Bernard, S. Descotes-Genon and L. Vale Silva, *Short-distance QCD corrections to* $K^0\bar{K}^0$ *mixing at next-to-leading order in left-right models*, JHEP **08** (2016) 128 [arXiv:1512.00543] [INSPIRE].

[43] P. Ball, J.M. Frere and J. Matias, *Anatomy of mixing induced CP asymmetries in left-right symmetric models with spontaneous CP-violation*, Nucl. Phys. B **572** (2000) 3 [hep-ph/9910211] [INSPIRE].

[44] I.I.Y. Bigi and J.M. Frere, *Strong radiative corrections to strangeness changing processes in the presence of right-handed currents*, Phys. Lett. B **129** (1983) 469 [*Erratum ibid.* **154** (1985) 457] [INSPIRE].

[45] G. Branco, J.-M. Frère and J.-M. Gérard, *The value of* $\epsilon'/\epsilon$ *in models based on* $SU(2)_L \times SU(2)_R \times U(1)$, Nucl. Phys. B **221** (1983) 317.

[46] K.S. Babu, K. Fujikawa and A. Yamada, *Constraints on left-right symmetric models from the process* $b \to s\gamma$, Phys. Lett. B **333** (1994) 196 [hep-ph/9312315] [INSPIRE].

[47] G. Bambhaniya, J. Chakrabortty, J. Gluza, T. Jeliński and M. Kordiaczynska, *Lowest limits on the doubly charged Higgs boson masses in the minimal left-right symmetric model*, Phys. Rev. D **90** (2014) 095003 [arXiv:1408.0774] [INSPIRE].

[48] A. Maiezza, M. Nemevšek and F. Nesti, *Perturbativity and mass scales in the minimal left-right symmetric model*, Phys. Rev. D **94** (2016) 035008 [arXiv:1603.00360] [INSPIRE].

[49] D. Guadagnoli and R.N. Mohapatra, *TeV scale left right symmetry and flavor changing neutral Higgs effects*, Phys. Lett. B **694** (2011) 386 [arXiv:1008.1074] [INSPIRE].

[50] R.N. Mohapatra and Y. Zhang, *LHC accessible second Higgs boson in the left-right model*, Phys. Rev. D **89** (2014) 055001 [arXiv:1401.0018] [INSPIRE].

[51] M. Frank, Ö. Özdal and P. Poulose, *Relaxing LHC constraints on the* $W_R$ *mass*, Phys. Rev. D **99** (2019) 035001 [arXiv:1812.05681] [INSPIRE].

[52] R.N. Mohapatra, G. Yan and Y. Zhang, *Ameliorating Higgs induced flavor constraints on TeV scale* $W_R$, Nucl. Phys. B **948** (2019) 114764 [arXiv:1902.08601] [INSPIRE].

[53] M. Frank, C. Majumdar, P. Poulose, S. Senapati and U.A. Yajnik, *Exploring* $0\nu\beta\beta$ *and leptogenesis in the alternative left-right model*, Phys. Rev. D **102** (2020) 075020 [arXiv:2008.12270] [INSPIRE].

[54] ATLAS collaboration, *Search for heavy Majorana or Dirac neutrinos and right-handed W gauge bosons in final states with two charged leptons and two jets at* $\sqrt{s} = 13$ *TeV with the ATLAS detector*, JHEP **01** (2019) 016 [arXiv:1809.11105] [INSPIRE].

[55] ATLAS collaboration, *Search for a right-handed gauge boson decaying into a high-momentum heavy neutrino and a charged lepton in pp collisions with the ATLAS detector at* $\sqrt{s} = 13$ *TeV*, Phys. Lett. B **798** (2019) 134942 [arXiv:1904.12679] [INSPIRE].

[56] CMS collaboration, *Search for heavy neutrinos and third-generation leptoquarks in hadronic states of two $\tau$ leptons and two jets in proton-proton collisions at $\sqrt{s} = 13$ TeV*, JHEP **03** (2019) 170 [arXiv:1811.00806] [INSPIRE].

[57] CMS collaboration, *Search for a heavy right-handed $W$ boson and a heavy neutrino in events with two same-flavor leptons and two jets at $\sqrt{s} = 13$ TeV*, JHEP **05** (2018) 148 [arXiv:1803.11116] [INSPIRE].

[58] Particle Data Group collaboration, *Review of particle physics*, Phys. Rev. D **98** (2018) 030001 [INSPIRE].

[59] ATLAS collaboration, *Search for doubly charged Higgs boson production in multi-lepton final states with the ATLAS detector using proton-proton collisions at $\sqrt{s} = 13$ TeV*, Eur. Phys. J. C **78** (2018) 199 [arXiv:1710.09748] [INSPIRE].

[60] R.N. Mohapatra, *Limits on the mass of the right-handed Majorana neutrino*, Phys. Rev. D **34** (1986) 909 [INSPIRE].

[61] J. Chakrabortty, J. Gluza, T. Jelinski and T. Srivastava, *Theoretical constraints on masses of heavy particles in left-right symmetric models*, Phys. Lett. B **759** (2016) 361 [arXiv:1604.06987] [INSPIRE].

[62] A. Maiezza, G. Senjanović and J.C. Vasquez, *Higgs sector of the minimal left-right symmetric theory*, Phys. Rev. D **95** (2017) 095004 [arXiv:1612.09146] [INSPIRE].

[63] G. Chauhan, P.S.B. Dev, R.N. Mohapatra and Y. Zhang, *Perturbativity constraints on $U(1)_{B-L}$ and left-right models and implications for heavy gauge boson searches*, JHEP **01** (2019) 208 [arXiv:1811.08789] [INSPIRE].

[64] I.Z. Rothstein, *Renormalization group analysis of the minimal left-right symmetric model*, Nucl. Phys. B **358** (1991) 181 [INSPIRE].

[65] J. Chakrabortty, P. Konar and T. Mondal, *Constraining a class of $B - L$ extended models from vacuum stability and perturbativity*, Phys. Rev. D **89** (2014) 056014 [arXiv:1308.1291] [INSPIRE].

[66] M.L. Swartz, *Limits on doubly charged Higgs bosons and lepton flavor violation*, Phys. Rev. D **40** (1989) 1521 [INSPIRE].

[67] M. Lusignoli and S. Petrarca, *Exotic Higgs production at $e^+e^-$ colliders*, Phys. Lett. B **226** (1989) 397 [INSPIRE].

[68] R.N. Mohapatra, *Rare decays of the tau lepton as a probe of the left-right symmetric theories of weak interactions*, Phys. Rev. D **46** (1992) 2990 [INSPIRE].

[69] SINDRUM collaboration, *Search for the decay $\mu^+ \to e^+e^+e^-$*, Nucl. Phys. B **299** (1988) 1 [INSPIRE].

[70] M. Doi, T. Kotani and E. Takasugi, *Double beta decay and Majorana neutrino*, Prog. Theor. Phys. Suppl. **83** (1985) 1 [INSPIRE].

[71] V. Tello, M. Nemevšek, F. Nesti, G. Senjanović and F. Vissani, *Left-right symmetry: from LHC to neutrinoless double beta decay*, Phys. Rev. Lett. **106** (2011) 151801 [arXiv:1011.3522] [INSPIRE].

[72] S.P. Das, F.F. Deppisch, O. Kittel and J.W.F. Valle, *Heavy neutrinos and lepton flavour violation in left-right symmetric models at the LHC*, Phys. Rev. D **86** (2012) 055006 [arXiv:1206.0256] [INSPIRE].

[73] J. Barry and W. Rodejohann, *Lepton number and flavour violation in TeV-scale left-right symmetric theories with large left-right mixing*, *JHEP* **09** (2013) 153 [arXiv:1303.6324] [INSPIRE].

[74] M. Drewes, *The phenomenology of right handed neutrinos*, *Int. J. Mod. Phys. E* **22** (2013) 1330019 [arXiv:1303.6912] [INSPIRE].

[75] F.F. Deppisch, P.S. Bhupal Dev and A. Pilaftsis, *Neutrinos and collider physics*, *New J. Phys.* **17** (2015) 075019 [arXiv:1502.06541] [INSPIRE].

[76] P.S.B. Dev, R.N. Mohapatra and Y. Zhang, *Probing the Higgs sector of the minimal left-right symmetric model at future hadron colliders*, *JHEP* **05** (2016) 174 [arXiv:1602.05947] [INSPIRE].

[77] T. Han and B. Zhang, *Signatures for Majorana neutrinos at hadron colliders*, *Phys. Rev. Lett.* **97** (2006) 171804 [hep-ph/0604064] [INSPIRE].

[78] S.R. Elliott and J. Engel, *Double beta decay*, *J. Phys. G* **30** (2004) R183 [hep-ph/0405078] [INSPIRE].

[79] KATRIN collaboration, *KATRIN design report* 2004, FZKA-7090, (2005) [INSPIRE].

[80] KamLAND-Zen collaboration, *Search for Majorana neutrinos near the inverted mass hierarchy region with KamLAND-Zen*, *Phys. Rev. Lett.* **117** (2016) 082503 [*Addendum ibid.* **117** (2016) 109903] [arXiv:1605.02889] [INSPIRE].

[81] EXO collaboration, *Search for neutrinoless double-beta decay with the upgraded EXO*-200 *detector*, *Phys. Rev. Lett.* **120** (2018) 072701 [arXiv:1707.08707] [INSPIRE].

[82] CUORE collaboration, *First results from CUORE: a search for lepton number violation via* $0\nu\beta\beta$ *decay of* $^{130}Te$, *Phys. Rev. Lett.* **120** (2018) 132501 [arXiv:1710.07988] [INSPIRE].

[83] CUPID-0 collaboration, *First result on the neutrinoless double-*$\beta$ *decay of* $^{82}Se$ *with CUPID*-0, *Phys. Rev. Lett.* **120** (2018) 232502 [arXiv:1802.07791] [INSPIRE].

[84] Majorana collaboration, *A search for neutrinoless double-beta decay in* $^{76}Ge$ *with* 26 *kg-yr of exposure from the Majorana demonstrator*, *Phys. Rev. C* **100** (2019) 025501 [arXiv:1902.02299] [INSPIRE].

[85] GERDA collaboration, *Probing Majorana neutrinos with double-*$\beta$ *decay*, *Science* **365** (2019) 1445 [arXiv:1909.02726] [INSPIRE].

[86] *NuFIT* 3.2 *webpage*, http://www.nu-fit.org, (2018).

[87] B. Kayser, F. Gibrat-Debu and F. Perrier, *The physics of massive neutrinos*, *World Sci. Lect. Notes Phys.* **25** (1989) 1.

[88] A. Belyaev, N.D. Christensen and A. Pukhov, *CalcHEP* 3.4 *for collider physics within and beyond the Standard Model*, *Comput. Phys. Commun.* **184** (2013) 1729 [arXiv:1207.6082] [INSPIRE].

[89] A. Roitgrund, G. Eilam and S. Bar-Shalom, *Implementation of the left-right symmetric model in FeynRules*, *Comput. Phys. Commun.* **203** (2016) 18 [arXiv:1401.3345] [INSPIRE].

[90] T. Sjöstrand, S. Mrenna and P.Z. Skands, *PYTHIA* 6.4 *physics and manual*, *JHEP* **05** (2006) 026 [hep-ph/0603175] [INSPIRE].

[91] T. Sjöstrand et al., *An introduction to PYTHIA* 8.2, *Comput. Phys. Commun.* **191** (2015) 159 [arXiv:1410.3012] [INSPIRE].

[92] R. Gavin, Y. Li, F. Petriello and S. Quackenbush, *W physics at the LHC with FEWZ* 2.1, *Comput. Phys. Commun.* **184** (2013) 208 [arXiv:1201.5896] [INSPIRE].

[93] G. Brooijmans et al., *Les Houches* 2013*: physics at TeV colliders. New physics working group report*, arXiv:1405.1617 [INSPIRE].

[94] DELPHES 3 collaboration, *DELPHES* 3*, a modular framework for fast simulation of a generic collider experiment*, *JHEP* **02** (2014) 057 [arXiv:1307.6346] [INSPIRE].

[95] J. Alwall et al., *The automated computation of tree-level and next-to-leading order differential cross sections, and their matching to parton shower simulations*, *JHEP* **07** (2014) 079 [arXiv:1405.0301] [INSPIRE].